\documentclass[3p,times,authoryear]{elsarticle}

\pdfoutput=1 
\usepackage[colorlinks=true,linkcolor=black, citecolor=blue, urlcolor=blue]{hyperref}
\usepackage{ecrc}
\usepackage{amsmath}

\usepackage[T1]{fontenc}
\usepackage{newtxtext,newtxmath}

\DeclareSymbolFont{gspecialletter}{U}{zeur}{}{n}
\DeclareMathSymbol{\specialg}{\mathord}{gspecialletter}{`g}
\DeclareMathAlphabet{\mathbbmsl}{U}{bbm}{m}{sl}

\volume{00}

\firstpage{1}

\journalname{International Journal of Engineering Science}

\runauth{Peruzzo et al.}

\jid{procs}

\jnltitlelogo{Int. J. Eng. Sci.}

\newcommand{\dispdot}[2][.2ex]{\dot{\raisebox{0pt}[\dimexpr\height+#1][\depth]{$#2$}}}

\usepackage[figuresright]{rotating}
\usepackage{multirow}

\begin{document}

\begin{frontmatter}



\dochead{}

\title{The Energy Balance of a Hydraulic Fracture at Depth\footnote{Submitted for review on July 8, 2024} }


\author{Carlo Peruzzo, Andreas M\"{o}ri, Brice Lecampion\footnote{Corresponding author: \texttt{Brice.Lecampion@epfl.ch}}}

\address{Geo-Energy Lab - Gaznat Chair, EPFL, station 18, CH-1015 Lausanne, Switzerland}

\begin{abstract}
We detail the energy balance of a propagating hydraulic fracture.
Using the linear hydraulic fracture model which combines lubrication flow and linear elastic fracture mechanics, we demonstrate how  different propagation regimes are related to the dominance of a given term of the power balance of a growing hydraulic fracture. Taking an energy point of view allows us to offer a physical explanation of  hydraulic fracture growth behaviours, such as, for example, the transition from viscosity to toughness dominated growth for a radial geometry, fracture propagation after the end of the injection or transition to self-buoyant elongated growth.
We quantify the evolution of the different power terms for a series of numerical examples. We also discuss the order of magnitudes  
of the different terms for a industrial-like hydraulic fracturing treatment accounting for the additional dissipation in the injection line.
\end{abstract}


\end{frontmatter}


\section{Introduction}

Hydraulic fractures are tensile mode I cracks that propagate 
in a pre-stressed solid due to an internal fluid pressure exerted on their opposite faces. Occurrences of hydraulic fracture growth range from magmatic intrusions in the upper Earth's crust  \citep{RiTa15} to the anthropogenic stimulation of wells in hydrocarbon reservoirs \citep{EcNo00} and pre-conditioning of rock masses in block caving mining \citep{HeSu16}. 

Over the years, the theoretical understanding of the growth of hydraulic fractures has significantly matured. The so-called linear hydraulic fracture mechanics model \citep{KhZh55,SpTu85,DeDe94,Deto16}  combines i) linear elastic fracture mechanics with ii) lubrication flow inside the fracture and iii) a simple fluid leak-off model. It can accurately reproduce controlled laboratory experiments measuring fracture growth under different conditions and in different materials \citep{BuDe08,BuGo13,LaZh15,LeDe17,TaSi23}.
Theoretical works have notably quantified the importance of the competition between the different dissipative processes at play during hydraulic fracture growth: fracture surface creation, viscous fluid losses, fluid leak-off and storage (see  \cite{Deto16} and references therein). 
Different limiting propagation regimes have been delineated usually using the strong form of the linear hydraulic fracture mechanics model. However, very few publications \citep{LeDe07,Bung13} have taken an energy balance point of view on the problem. 

In this paper, we develop the energy balance for linear hydraulic fracture mechanics by combining the balance for the elastic solid and the fracturing fluid. In doing so, we highlight the consistency of the different assumptions behind the linear hydraulic fracture model. The main benefit of this energy approach is to clearly demonstrate the competition between the different processes at play during hydraulic fracture growth and the resulting scaling laws.

For the solid and fluid systems, taken separately, the energy equation is given by the balance of the change of the internal $\textrm{U}$ and kinetic $\textrm{K}$ energies with the heat variation $\text{dQ}$ and the work of the external forces $\text{d} \textrm{W}_{\textrm{ext}}$:
\begin{equation}
    \text{d} \textrm{U} + \text{d} \textrm{K} = \text{dQ} + \text{d} \textrm{W}_{\textrm{ext}}. 
    \label{eq:E_conservation_general_U}
\end{equation}

The second principle of thermodynamics specifies that the change of entropy of the system $\textrm{S}$ must be greater or equal to the heat variation over the temperature  $\textrm{T}$: $\textrm{dS} \ge \text{dQ}  /\textrm{T}  $.
%
Introducing an energy dissipation $D$, the second principle can be rewritten as $\textrm{T} \textrm{dS} = \text{dQ}  + \text{d} D$, with $ \text{d} D\ge 0$. In this expression, the dissipation term $\text{d} D$ resolves the inequality and highlights the non-reversibility of the process. 
Introducing the Helmholtz free energy $\Psi = \textrm{U} - \textrm{TS} $, under isothermal condition $\text{d} \textrm{T} = 0$ (assumed throughout this entire contribution for simplicity), the energy equation can thus be expressed as
\begin{equation}
    \text{d}\Psi + \text{d} \textrm{K} + \text{d} D= \text{d} \textrm{W}_{\textrm{ext}}, 
    \label{eq:E_conservation_general}
\end{equation}
with the constraint that the rate of dissipation is either positive or null, $ \text{d} D\ge 0$.

In what follows, we particularize the energy equation first for the solid, and then for the fracturing fluid. The energy dissipation will be associated with fracture surface creation in the solid and viscous flow in the fluid.
We then discuss the conditions at the fluid-solid interface, notably in view of the exchange of fluid between the fracture and the permeable solid.  We finally arrive at a global energy balance for the linear hydraulic fracture mechanics model by combining the fluid and solid energy balance. We subsequently show that the dimensionless numbers obtained from scaling arguments such as the dimensionless toughness or dimensionless storage coefficients are indeed directly related to ratios of powers. We showcase the evolution of the different power terms on a number of examples for planar hydraulic fractures in infinite media. We then estimate the additional energy losses in fluid flow in the injection line (e.g. wellbore in practical applications) and discuss orders of magnitude found in practice.

\section{The power balance of a pre-stressed solid undergoing quasi-static fracture growth\label{sec:solid}}

We consider the case of a hydraulically fractured solid of domain $\Omega^\textrm{S}$ and boundary $\partial \Omega^\textrm{S}$ initially (in the reference configuration) under a state of pre-existing compressive stress $\sigma_{ij}^o$. Such an initial stress field $\sigma_{ij}^o$  must be in equilibrium with the solid body forces from gravity as well as any far-field tectonic strain. Under almost all geological configurations, $\sigma_{ij}^o$ increases linearly with depth, with one of the principal stress directions aligned with the gravitational vector. The solid mechanical deformation under the assumption of a linear reversible elastic behaviour is thus governed by the increment of stress over this initial stress field $\sigma_{ij}-\sigma_{ij}^o$, which must be in equilibrium without any acting body forces (which are accounted for in the initial stress field). 

We briefly recap below well-known results for the energy balance in linear elastic fracture mechanics for clarity (see \cite{Rice68,Rice78,KeSi96,Lebl03,Bui06} among many others for a detailed presentation). Under isothermal conditions and for a linear elastic solid, the Helmholtz free energy $\Psi$ reduces to the elastic strain energy:
\begin{equation}
    \Psi^\textrm{S} = \int_{\Omega^\textrm{S}}{\frac{1}{2}\,\epsilon_{ij}\,(\sigma_{ij}-\sigma_{ij}^o)\,\textrm{d}V,}
\end{equation}
where $\Psi^S$ accounts for the pre-stress $\sigma_{ij}^o$ and $\epsilon_{ij}$ is the strain tensor.
Particularizing the conservation of energy \eqref{eq:E_conservation_general} for the elastic solid (with a superscript $\textrm{S}$), under the assumption of quasi-static fracture growth (e.g., negligible variation of kinetic energy of the solid,  $\textrm{dK}^\textrm{S}=0$), the variation of external work is balanced by the variation of elastic energy and the rate of energy dissipated in the creation of new fracture surfaces $\textrm{d}D^\textrm{S}\ge 0 $:
\begin{equation}
\textrm{d}\textrm{W}_{\textrm{ext}}^\textrm{S}=\textrm{d}\Psi^\textrm{S}+
\textrm{d}D^\textrm{S}. \label{eq:en_cons_S}
\end{equation} 
The variation of the external work $\textrm{d} \textrm{W}_{\textrm{ext}}^{\textrm{S}}$ for a pre-stressed solid is given by the work of external tractions $T_i$ applied over the boundary $\partial\Omega^{\textrm{S}}_T$ in a displacement increment $\textrm{d} u_i$: 
\begin{equation}
\label{E_extS}
\textrm{d} \textrm{W}_{\textrm{ext}}^{\textrm{S}}=\int_{\partial\Omega^{\textrm{S}}_T}{(T_i-T_i^o)\, \textrm{d} u_i\,\textrm{d}S} ,\qquad i=1,2,3,
\end{equation}
where $T_i^o=\sigma_{ij}^on_j$ denotes the tractions corresponding to the initial stress field (which include the effect of gravitational body forces), with $n_j$ the outward unit normal to the boundary. 


It is important to realize that the total variation $\textrm{d}\left(\cdot\right)$ of each energy term $\textrm{W}_{\textrm{ext}}^\textrm{S}$, $\Psi^\textrm{S}$, 
or $D^\textrm{S}$ in Eq.~\eqref{eq:en_cons_S} can be caused, either by a variation of external forces $\mathcal{F}$ (or imposed displacements), or/and an increment of fracture area $a$. We refer to $\mathcal{F}$ as $\textit{the generalized load}$, intending any combination of external force and applied displacements. 
The generalized load $\mathcal{F}$ and the fracture area $a$ are independent variables, allowing us to express the derivative $\textrm{d}\left(\cdot\right)$ applied to each energy term of  Eq.~\eqref{eq:en_cons_S}, as 
\begin{equation}
\label{eq:totalD}
\textrm{d}\left(\cdot\right)=\left.\frac{\partial\left(\cdot\right)}{\partial\mathcal{F}}\right|_{a=\textrm{const.}}\textrm{d}\mathcal{F}+\left.\frac{\partial\left(\cdot\right)}{\partial a}\right|_{\mathcal{F}=\textrm{const.}}\textrm{d}a,
\end{equation}
where the notations $\left(\cdot\right)|_{a=\textrm{const}}$, and $\left(\cdot\right)|_{\mathcal{F}=\textrm{const}}$ emphasize that the derivative is taken with respect to a constant fracture area $a$ and a constant generalized load $\mathcal{F}$.

By definition, a variation of the energy dissipated in the creation of new surfaces 
only occurs if additional fracture area is created during the process ($\textrm{d}a \ge 0$):
\begin{equation}\label{eq:defdDdF}
\left.\frac{\partial D^\textrm{S}}{\partial\mathcal{F}}\right|_{a=\textrm{const.}}=0.
\end{equation}
The total variation $\textrm{d}D^\textrm{S}$  is expressed as the integral, along the fracture front $\Gamma$, of the local critical fracture energy $G_c$ (a material property) times the local fracture advancement $\textrm{d}\ell  \textrm{d}\gamma $ 
\begin{equation}\label{eq:defG}
\textrm{d}D^\textrm{S} = \left.\frac{\partial D^\textrm{S}}{\partial a}\right|_{\mathcal{F}= \textrm{const.}} \textrm{d}a
= \int_{\Gamma} G_c   \textrm{d}\ell  \textrm{d}\gamma, 
\end{equation}
with $\gamma$ denoting the curvilinear coordinate along the fracture front, and $\textrm{d}\ell$ the fracture increment in the direction normal to the fracture front.
By virtue of the principle of virtual powers applied to a linear elastic material we additionally obtain,
\begin{equation}
\label{eq:pvp}
\left.\frac{\partial\left(\textrm{W}_{\textrm{ext}}^\textrm{S}-\Psi^\textrm{S}\right)}{\partial\mathcal{F}}\right|_{a=\textrm{const.}}=0,
\end{equation}
such that Eq. \eqref{eq:en_cons_S}, with (\ref{eq:defG}), reduces to the equivalence
\begin{equation}
\label{eq:GeqGc}
\mathbbmsl{G}\,\textrm{d}a=\textrm{d}D^\textrm{S}=\int_{\Gamma} G_c\,\textrm{d}\ell\, \textrm{d}\gamma,
\end{equation}
in which $\mathbbmsl{G}$ is the classical definition of the energy release rate for quasi-static propagation (which implies $\textrm{dK}^\textrm{S}=0$)
\begin{equation}
\mathbbmsl{G}\coloneqq\left.\frac{\partial\left(\textrm{W}_{\textrm{ext}}^\textrm{S}-\Psi^\textrm{S}\right)}{\partial a}\right|_{\mathcal{F}=\textrm{const.}},\label{eq:G_1}
\end{equation}
also often written as 
\begin{equation}
\mathbbmsl{G}\coloneqq-\left.\frac{\partial\phi}{\partial a}\right|_{\mathcal{F}=\textrm{const.}},\label{eq:G_2}
\end{equation}
where $\phi=\Psi^\textrm{S}-\textrm{W}_{\textrm{ext}}^\textrm{S}$ is the total elastic potential energy \citep{Ri68}.
In~\ref{appendix:deriv_in_integral}, we also show that the classical expression of the energy release rate \citep{KeSi96}  \begin{equation}\label{eq:G_def}
\mathbbmsl{G}=\frac{1}{2}\int_{\partial\Omega^\textrm{S}}\left(\frac{\mathrm{\mathrm{\partial}}u_{i}}{\mathrm{\mathrm{\partial}}a}(T_{i}-T_{i}^o)-u_{i}\frac{\mathrm{\mathrm{\partial}}(T_{i}-T_{i}^o)}{\mathrm{\mathrm{\partial}}a}\right)\mathrm{d}S,
\end{equation}
remains valid when the fracture surfaces are loaded. 


%
We map the evolution of external forces in time $t$, such that $\textrm{d}\mathcal{F}=\textrm{d}t$, and use the fact that the derivatives $\partial(\cdot)/\partial a$ with respect to the crack area $a$ in Eq.~\eqref{eq:G_def} are linked to the time derivative $\textrm{d}(\cdot)/\textrm{d}t$ as:
\begin{equation}\label{eq:timederiv}
\frac{\mathrm{\mathrm{\partial}}\left(\cdot\right)}{\mathrm{\mathrm{\partial}}a}=\frac{\partial\left(\cdot\right)}{\partial t}\left(\frac{\mathrm{d}a}{\mathrm{d}t}\right)^{-1},
\end{equation}
where $\textrm{d}a/\textrm{d}t$ is the rate of growth of the fracture area $a$. The energy conservation in Eq.~\eqref{eq:en_cons_S} in combination with Eqs.~\eqref{eq:GeqGc},  \eqref{eq:G_def}, and \eqref{eq:timederiv} for the case of negligible inertia ($\textrm{dK}^\textrm{S}=0$) can be re-written as:
\begin{equation}
\underset{\textrm{External\,power}}{\underbrace{\int_{\partial\Omega^{\textrm{S}}}\frac{\partial {u}_i}{\partial t}\,(T_i-T_i^o)\,\mathrm{d}S}}=\underset{\textrm{Internal\,power}}{\underbrace{\underset{\textrm{Storage\,rate\,of\,elastic\,energy}}{\underbrace{\int_{\partial\Omega^{\textrm{S}}}\frac{\partial}{\partial t}\left[\frac{1}{2}u_i(T_i-T_i^o)\right]\mathrm{d}S}}+\underset{\textrm{Dissipation in creation of frac. surfaces}}{\underbrace{\int_{\Gamma}G_{\textrm{c}}\,v_i\,n_i\,\mathrm{d}\gamma}}}},\label{eq:interm_balance_S}
\end{equation}
where $G_c$ is the local critical fracture energy and $v_i  n_i=\textrm{d}\ell/\textrm{d}t$  the local velocity of the fracture front $\Gamma$. Eq.~\eqref{eq:interm_balance_S} states that the total external power exerted by the external forces is balanced by the sum of the rate of dissipated energy via fracture advancement and the rate of energy elastically stored in the medium. 

Finally, the solid energy balance Eq.~\eqref{eq:interm_balance_S} can be further written by highlighting the contribution of the external power from the fluid-solid interfaces $\partial\Omega^{\textrm{S}\leftrightarrow \textrm{F}}=\Sigma^+ \bigcup \Sigma^-$ (the union of the top and bottom surfaces of the fracture) and the external solid boundaries $\partial\Omega^{\textrm{S}}_\Box$:
\begin{equation}
\int_{\partial\Omega^{\textrm{S}\leftrightarrow \textrm{F}}}\frac{\partial u_i}{\partial t}\,(T_i-T_i^o)\,\mathrm{d}S+\int_{\partial\Omega^{\textrm{S}}_\Box}\frac{\partial u_i}{\partial t}\,(T_i-T_i^o)\,\mathrm{d}S=\int_{\partial\Omega^{\textrm{S}}}\frac{\partial}{\partial t}\left[\frac{1}{2}u_i\,\left(T_i-T_i^\textrm{o}\right)\right]\mathrm{d}S+\int_{\Gamma}G_\textrm{c}\,v_i\,n_i\,\mathrm{d}\gamma.
\label{eq:Solid_balance_final}
\end{equation}
The contribution from the external solid boundary $\partial\Omega^{\textrm{S}}_\Box$ is typically absent for problems at depth where the solid medium can be considered as infinite but arise, for example, when applied tractions or displacement are modified during an experiment on a finite size specimen. We drop this term associated with the external solid boundary in the remainder of this paper for simplicity.

The global fracture energy balance is often written locally at any point along the fracture front $\Gamma$ as the following {\it local} propagation criteria, highlighting the irreversibility of fracture growth:
\begin{equation}
    \begin{cases}
      \,\left(G-G_\textrm{c}\right)v_i n_i\ge 0, & \\
      \, G_c v_i n_i\ge 0, &
    \end{cases}
\end{equation}
which states that, if the fracture is propagating ($v_in_i\neq0$), the local energy release rate $G$ must equal the local material fracture energy $G_\textrm{c}$. 

\begin{figure}[]
\noindent \begin{centering}
\includegraphics[width=0.9\textwidth]{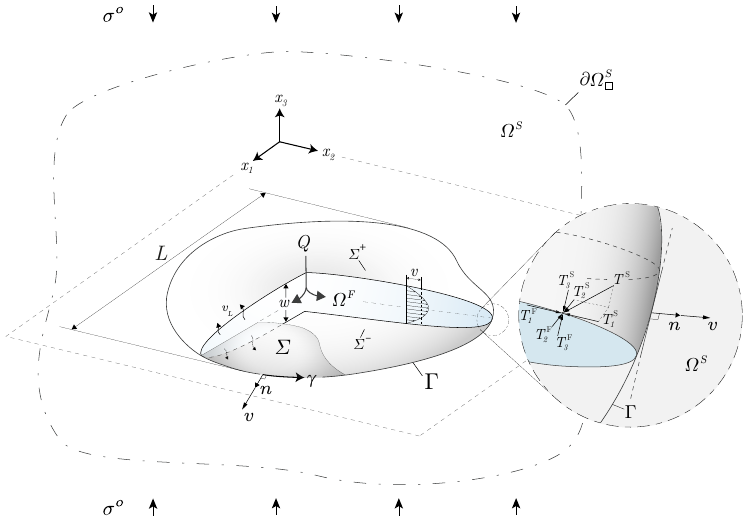}
\par\end{centering}
\caption{Scheme of a hydraulic fracture buried at depth. $\Sigma$ is the middle plane between the upper $\Sigma^+$  and lower $\Sigma^-$ solid-fluid interface. The fluid velocity $v$ coincides with the fracture front velocity along the fracture front in the absence of a fluid lag \citep{DePe14}. \label{fig:scheme_3D}}
\end{figure}


\section{The power balance of fluid flow in the lubrication approximation}
\subsection{Preliminaries}
The conservation of mechanical energy Eq.~\eqref{eq:E_conservation_general} for a  fluid of volume $\Omega^\textrm{F}$ undergoing a isothermal transformation can be expressed as
\begin{equation}
\dispdot{\textrm{W}}_{\textrm{ext}}^\textrm{F}=\dispdot{\Psi}^\textrm{F}+\dispdot{\textrm{K}}^\textrm{F}+\dispdot{D}^\textrm{F}, \label{eq:en_cons_F}
\end{equation} 
where the dot " $\dot{\textrm{ }}$ " over a quantity represents the derivative with respect to time. In passing from  Eq.~\eqref{eq:E_conservation_general} to Eq.~\eqref{eq:en_cons_F} we have used that the variation $\textrm{ d}\left(\cdot\right)$ of each energy term $\textrm{W}_{\textrm{ext}}^\textrm{F}$, $\Psi^\textrm{F}$, $\textrm{K}^\textrm{F}$, or $D^\textrm{F}$ can be directly mapped to time. The isothermal power balance for the fluid \eqref{eq:en_cons_F} states that the rate of work $\dispdot{\textrm{W}}_{\textrm{ext}}^\textrm{F}$ 
\begin{equation}
\dispdot{\textrm{W}}_{\textrm{ext}}^\textrm{F}=\int_{\partial\Omega^{\textrm{F}}_T}{T_i\,\dot{u}_i\,\textrm{d}S}+\int_{\Omega^\textrm{F}}{\rho^\textrm{F}\,g_i\,\dot{u}_i\,\textrm{d}V},\qquad i=1,2,3
\end{equation}
done to the fluid volume by the external traction $T_i$, applied over the boundary $\partial\Omega^{\textrm{F}}_T$, and by the presence of body forces $\left(\rho^\textrm{F}\,g_i\right)$ (where $\rho^\textrm{F}$ is the fluid density, and $g_i$ is the gravity vector), is balanced by the sum of the rates of Helmholtz free energy $\dispdot{\Psi}^F$, kinetic energy
\begin{equation}
\dispdot{\textrm{K}}^\textrm{F}=\int_{\Omega^{\textrm{F}}}\frac{\textrm{D}}{\textrm{D}t}\left(\frac{1}{2}\,\rho^\textrm{F}\,\dot{u}_{i}^{2}\right)\textrm{d}V.
\end{equation}
and dissipation $\dispdot{D}^\textrm{F}$. In order to identify explicitly the rate of viscous dissipation 
$\dispdot{D}^\textrm{F}$, one can first rewrite the conservation of energy using the kinetic energy theorem as 
\citep{Batc67}:
\begin{equation}
   \dispdot{\textrm{W}}_{\textrm{ext}}^\textrm{F}
   = 
    \mathcal{P}^F_{int}(\dot{u}_i)+
    \dispdot{\textrm{K}}^\textrm{F},
    \label{eq:energy_with_kt}
\end{equation}
where $\mathcal{P}^F_{int}(\dot{u}_i)$ is the power of internal forces
\begin{equation}
    \mathcal{P}^F_{int}(\dot{u}_i) =\int_{\Omega^\textrm{F}} 
    \sigma_{ij}  d_{ij}\textrm{d} V ,\qquad i,j=1,2,3
\end{equation}
 $d_{ij}$ being the strain rate tensor $d_{ij} = \frac{1}{2}\left( \dfrac{\partial \dot{u}_{i}}{\partial x_j} + \dfrac{\partial \dot{u}_j}{\partial x_i}\right)$. 
For an isothermal transformation, the variation of Helmholtz free energy of a fluid is simply 
\begin{equation}
     \dispdot{\Psi}^\textrm{F}= -\int_{\Omega^\textrm{F}} p \dfrac{\partial \dot{u}_{i}}{\partial x_i}  \textrm{d} V, 
\end{equation}
where $p=-\sigma_{ii}/3$ is the fluid pressure, such that the rate of dissipation in the fluid can be identified from Eqs \eqref{eq:en_cons_F} and \eqref{eq:energy_with_kt} as $ \dispdot{D}^\textrm{F}= \mathcal{P}^F_{int}(\dot{u}_i) - \dispdot{\Psi}^\textrm{F}$.
For a Newtonian fluid, the stress tensor reads
\begin{equation}
    \sigma_{ij}=-p \delta_{ij} + 2 \mu \left(d_{ij} - \frac{1}{3} \frac{\partial \dot{u}_k}{\partial x_k} \delta_{ij}\right) 
    + \kappa  \frac{\partial \dot{u}_k}{\partial x_k} \delta_{ij},\qquad i,j,k=1,2,3, 
\end{equation}
where $ \delta_{ij} $ is the Kronecker delta, and $\mu$ and $\kappa$ are the shear and expansion viscosities. The rate of dissipation $\dispdot{D}^\textrm{F}$ is thus obtained as
\begin{equation}
\dispdot{D}^\textrm{F}=\underset{\textrm{dissipation\,by\,isotropic\,expansion/contraction}}{\underbrace{ \left(\kappa - 2/3 \mu \right)\int_{\Omega^{\textrm{F}}} \left(\frac{\partial \dot{u}_{k}}{\partial x_{k}}\right)^2\,\textrm{d}V}} + \underset{\mathrm{dissipation\,by\,shear}}{\underbrace{2\mu\int_{\Omega^{\textrm{F}}}d_{ij}d_{ij}\,\textrm{d}V}}.
\label{eq:D_F}
\end{equation}
Restricting to fracturing liquids which are only very slightly compressible $\left|\dfrac{\partial \dot{u}_{k}}{\partial x_{k}}\right|\ll 1$, the dissipation by expansion/contraction is negligible compared to the dissipation by shear (see \cite{Batc67} for discussion). 


\subsection{Thin film flow}

\begin{figure}[ht]
\noindent \begin{centering}
\includegraphics[width=0.5\textwidth]{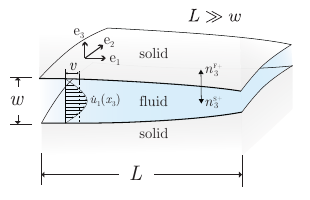}
\par\end{centering}
\caption{
The characteristic fracture extent $L$ is much larger than fracture opening $w$ ensuring validity of lubrication flow in the fracture. \label{fig:FL}}
\end{figure}

Although the discussion remains valid for a smoothly curved thin-film fluid flow \citep{szeri10}, we particularize here the fluid domain as consisting of a planar fluid-filled fracture, with its normal given by $\mathbf{e}_3$ (see Fig.~\ref{fig:FL} for a sketch). The width $w=u^{S+}_3-u^{S-}_3$ of a three-dimensional hydraulic fracture is much smaller than the characteristic dimension of its surface $L$ such that the ratio $\epsilon=w/L$ is much smaller than 1. This also implies that, if the components of the fluid velocity parallel to the crack faces (taken here as $\dot{u}_1$ and $\dot{u}_2$) scale as the characteristic velocity $\dot{U}$, then the component of the fluid velocity across the fracture opening ($\dot{u}_3$) scales as $\epsilon\dot{U}$.
In other words, we have $\epsilon\sim \dot{u}_1/\dot{u}_3  \sim  \dot{u}_2/\dot{u}_3\ll1$ in the coordinates system of Fig.~\ref{fig:FL}.  As a consequence, the equations governing the fluid flow are re-scaled and can be simplified by neglecting terms proportional to $\epsilon$. In particular, in this limit known as the "thin film approximation" (see for example \cite{szeri10}), the scaling of the pressure-driven problem is defined as:
\[
\left( \bar{x}_1, \; \bar{x}_2, \; \bar{x}_3 \right) = \left(\frac{x_1}{L},\;\frac{x_2}{L},\;\frac{x_3}{\epsilon\,L}\right),\qquad \left( \bar{\dot{u}}_1, \; \bar{\dot{u}}_2, \; \bar{\dot{u}}_3 \right) = \left(\frac{\dot{u}_1}{\dot{U}},\;\frac{\dot{u}_2}{\dot{U}},\;\frac{\dot{u}_3}{\epsilon\,\dot{U}}\right),
\]
\begin{equation}
p=\frac{\mu\,\dot{U}}{L\,\epsilon^2}\,\bar{p}, \quad \left( T_1,\,T_2,\,T_3\right)=\frac{\mu\,\dot{U}}{L\,\epsilon^2}\left(\epsilon \bar{T}_1, \epsilon \bar{T}_2, \bar{T}_3\right)\quad t=\frac{L}{\dot{U}}\,\bar{t},\quad g_i=\textrm{g}\,\bar{g}_i,\qquad i=1,2,3 \label{eq:lubrication_scaling}
\end{equation}
where we have put a bar " $\bar{}$ " above each symbol to denote the corresponding dimensionless quantity. In such thin-film scaling, the scaled version of the  power balance \eqref{eq:en_cons_F} can be expressed as
\begin{equation}
\begin{split}
\bar{\dispdot{\textrm{W}}}_{\textrm{ext}-\partial\Omega^{\textrm{F}}_T}^\textrm{F}+\frac{w^2\,\rho_\textrm{F}\,\textrm{g}}{\dot{U}\,\mu}\, \bar{\dispdot{\textrm{W}}}_{\textrm{ext}-\Omega^\textrm{F}}^\textrm{F} &=\bar{\dispdot{\Psi}}^\textrm{F}+ \epsilon \,\textrm{Re} \,\bar{\dispdot{\textrm{K}}}^\textrm{F} +\epsilon^2\,\left(\frac{\kappa}{\mu}-2/3 \right)\bar{\dispdot{\textrm{D}}}_{\textrm{iso}}^\textrm{F}+\bar{\dispdot{\textrm{D}}}_{\textrm{dev}-0}^\textrm{F} +\epsilon^2\,\bar{\dispdot{\textrm{D}}}_{\textrm{dev}-2}^\textrm{F}
+\epsilon^4\,\bar{\dispdot{\textrm{D}}}_{\textrm{dev}-4}^\textrm{F},
\end{split}
\end{equation}
where the $\textit{Reynolds}$ $\textit{number}$ $\textrm{Re}$ is defined as
\begin{equation}
\textrm{Re}=\frac{\rho_\textrm{F}\,w\,\dot{U}}{\mu},
\end{equation}
and the expressions for the different dimensionless powers
\begin{equation}
\bar{\dispdot{\textrm{W}}}_{\textrm{ext}-\partial\Omega^{\textrm{F}}_T}^\textrm{F}=\int_{\partial\Omega^{\textrm{F}}_T}{ \bar{T}_i\,\bar{\dot{u}}_i\textrm{d}\overline{S}}, \qquad
\bar{\dispdot{\textrm{W}}}_{\textrm{ext}-\Omega^\textrm{F}}^\textrm{F}=\int_{\Omega^\textrm{F}}{\left(\bar{g}_1\bar{\dot{u}}_1+\bar{g}_2\bar{\dot{u}}_2+\epsilon\,\bar{g}_3\bar{\dot{u}}_3\right)\textrm{d}\overline{V}},  
\end{equation} 
\begin{equation}
\bar{\dispdot{\Psi}}^\textrm{F}=\int_{\Omega^{\textrm{F}}}\bar{p}\frac{\partial\bar{\dot{u}}_{i}}{\partial \bar{x}_{i}}\textrm{d}\overline{V},
\end{equation} 
\begin{equation}
\bar{\dispdot{\textrm{K}}}^\textrm{F}=\int_{\Omega^{\textrm{F}}}\frac{1}{2}\,\frac{\textrm{D}}{\textrm{D}\bar{t}}\,\left(\,\bar{\dot{u}}_{1}^{2}+\bar{\dot{u}}_{2}^{2}+\epsilon^2\,\bar{\dot{u}}_{3}^{2}\,\right)\textrm{d}\overline{V},
\end{equation} 
\[
\bar{\dispdot{\textrm{D}}}_{\textrm{iso}}^\textrm{F}=\int_{\Omega^{\textrm{F}}} \frac{\partial \bar{\dot{u}}_{i}}{\partial \bar{x}_{i}}\,\frac{\partial \bar{\dot{u}}_{j}}{\partial \bar{x}_{j}}\,\textrm{d}\overline{V}, \qquad \bar{\dispdot{\textrm{D}}}_{\textrm{dev}-0}^\textrm{F}=\int_{\Omega^{\textrm{F}}}\left[\left(\frac{\partial \bar{\dot{u}}_{1}}{\partial \bar{x}_{3}}\right)^2+\left(\frac{\partial \bar{\dot{u}}_{2}}{\partial \bar{x}_{3}}\right)^2\right]\,\textrm{d}\overline{V},
\]
\begin{equation}
\bar{\dispdot{\textrm{D}}}_{\textrm{dev}-2}^\textrm{F}=\int_{\Omega^{\textrm{F}}}\left[\left(\frac{\partial \bar{\dot{u}}_{1}}{\partial \bar{x}_{2}}+\frac{\partial \bar{\dot{u}}_{2}}{\partial \bar{x}_{1}}\right)^2+2\left(\frac{\partial \bar{\dot{u}}_{2}}{\partial \bar{x}_{3}}\frac{\partial \bar{\dot{u}}_{3}}{\partial \bar{x}_{2}}+\frac{\partial \bar{\dot{u}}_{1}}{\partial \bar{x}_{3}}\frac{\partial \bar{\dot{u}}_{3}}{\partial \bar{x}_{1}}\right)\right]\,\textrm{d}\overline{V}, \qquad \bar{\dispdot{\textrm{D}}}_{\textrm{dev}-4}^\textrm{F}=\int_{\Omega^{\textrm{F}}}\left[\left(\frac{\partial \bar{\dot{u}}_{3}}{\partial \bar{x}_{1}}\right)^2+\left(\frac{\partial \bar{\dot{u}}_{3}}{\partial \bar{x}_{2}}\right)^2\right]\,\textrm{d}\overline{V},
\end{equation} 
are obtained by scaling the infinitesimal  volume element $\textrm{d}V$ as $\textrm{d}V=\epsilon L^3\,\textrm{d}\overline{V}$, and the surface element $\textrm{d}S$ as $\textrm{d}S=L^2\,\textrm{d}\overline{S}$. 

If we neglect all terms that are of order $\ge\mathcal{O}\left(\epsilon\right)$, 
we obtain the following power balance in dimensional form
\begin{equation}
\int_{\partial\Omega^\textrm{F}_T}{T_i\,\dot{u}_i\,\textrm{d}S}+\underset{\dispdot{\textrm{W}}_{\textrm{ext}-\Omega^\textrm{F}}^\textrm{F}}{\underbrace{\int_{\Omega^\textrm{F}} {\rho_{\textrm{F}}\,g_\alpha\,\dot{u}_\alpha\,\textrm{d}V}}}=\underset{\dispdot{\textrm{D}}_{\textrm{dev}-0}^\textrm{F}}{\underbrace{\int_{\Omega^{\textrm{F}}}\mu\left[\left(\frac{\partial\dot{u}_{2}}{\partial x_{3}}\right)^{2}+\left(\frac{\partial\dot{u}_{1}}{\partial x_{3}}\right)^{2}\right]\textrm{d}V}},\quad \alpha=1,2\quad i=1,2,3.\label{eq:Fluid_P_lubrication}
\end{equation}
It reveals that the external power is dissipated primarily because of the velocity gradient perpendicular to the flow direction. Although it has been written here for a Newtonian fluid, it can be easily extended to more complex rheology. It is also worth noting that the energy dissipated by inertia is of order $\epsilon\,\textrm{Re}$, hinting at the fact that it is negligible in fracture flow. Recently, other studies have demonstrated that the energy dissipated by inertia is indeed negligible under most practical circumstances \citep{ZiLe17,LeZi19,GeGr24}.

The external power, exchanged at the fluid boundary $\partial\Omega^{\textrm{F}}$, can be split into the one exchanged at the fluid-solid interface which consists of the union of the top and bottom surfaces of the fracture $ \partial\Omega^{\textrm{S}\leftrightarrow \textrm{F}}=\Sigma^+ \bigcup \Sigma^-$
and the one exchanged via an external fluid boundary $\partial\Omega^{\textrm{F}}_\textrm{o}$ which corresponds to the fluid surface that crosses the location of the wellbore where the fluid is injected. For a wellbore diameter that is much smaller compared to the fracture size, the power exchanged through the wellbore can be simply expressed as the product between the volumetric rate of fluid entering the fracture $Q_{in}$ and the corresponding fluid pressure $p_\textrm{in}$ at that point, such that:
\begin{equation}
\int_{\partial\Omega^\textrm{F}_T}{T_i\,\dot{u}_i\,\textrm{d}S}
=\int_{\partial\Omega^{\textrm{S}\leftrightarrow \textrm{F}}}{T_i\,\dot{u}_i\,\textrm{d}S}+Q_{in}\,p_{in},\quad i=1,2,3.\label{eq:ext_P_F}
\end{equation}

\section{The conditions at the fluid-solid interface}

The action-reaction law (Newton's third law) applies at the fluid-solid interfaces:
\begin{equation}
    T_i^{\textrm{S}}+ T_i^{\textrm{F}} = 0,\quad i=1,2,3.
\end{equation}
where $T_i^\textrm{S}$ denotes the traction applied to the solid and $T_i^{\textrm{F}}$ to the fluid respectively. This is valid at both fluid-solid interfaces: the top $\Sigma^+$ and bottom $\Sigma^-$ surfaces of the fracture. However, the fluid and solid velocities are not necessarily equal at the fluid-solid interface. 

In the case of porous rock, the injected fluid can leak off due to the difference between the fluid pressure in the fracture and the far-field pore pressure. This phenomenon is characterized by the fact that the fluid velocity in the direction normal to the fracture surfaces (direction $\mathbf{e}_3$ in our local system) is larger than the velocity with which the fracture width 
varies (the solid velocity), leading to a relative velocity between the fluid and solid phases in the direction normal to the fracture plane:
\begin{equation}
    \dot{u}_i^{\textrm{F}}-    \dot{u}_i^{\textrm{S}}
    = v_L \delta_{i3},\quad i=1,2,3,
\end{equation}
where $v_L$ is the so-called fluid leak-off velocity. We assume, however, a no-slip boundary condition for the velocity components in the fracture plane tangent to the fluid-solid interfaces: $\dot{u}_\alpha^{\textrm{F}}=   \dot{u}_\alpha^{\textrm{S}},\,\alpha=1,2$.

Locally, at any point along the fluid-solid interfaces in the fracture plane,  the solid and fluid power (per unit of surface) reads:
\begin{equation}
    T_i^{\textrm{S}} \dot{u}_i^{\textrm{S}} + T_i^{\textrm{F}} \dot{u}_i^{\textrm{F}} = T_3^{\textrm{F}}   v_L, \quad i=1,2,3.
    \label{eq:F_S_interface_power}
\end{equation}
Fluid leak-off therefore acts as an energy sink at the level of the fluid-solid interface. 

The interface condition \eqref{eq:F_S_interface_power} provides the closure relation needed to obtain the energy balance for the full system summing up the fluid and solid contributions. Before doing so, we further discuss the fluid tractions transmitted to the solid, and the intensity of the horizontal component of the power exchange $T_\alpha^{\textrm{F}} \dot{u}_\alpha^{\textrm{F}}$ ($\alpha=1,2$).

\subsection{The traction transmitted to the solid at the top and bottom interfaces}
We now decipher the traction at the top and bottom interfaces, with subscript $+$ and $-$ respectively. From the fluid side $ T_i^{\textrm{F}}=- T_i^{\textrm{S}}$, they are given by :
\begin{eqnarray}
    T_i^{\textrm{F}+} = \sigma_{ij}^{\textrm{F}}\left(x_3=\frac{w}{2}\right) n_j^{\textrm{F}+}= -p \delta_{i3} + \sigma_{\alpha 3}\left(x_3=\frac{w}{2}\right) \delta_{i \alpha},\quad \alpha=1,2, \\
    T_i^{\textrm{F}-}= \sigma_{ij}^{\textrm{F}} \left(x_3=-\frac{w}{2}\right) n_j^{\textrm{F}-}= p \delta_{i3} - \sigma_{\alpha 3}\left(x_3=-\frac{w}{2}\right) \delta_{i \alpha},\quad \alpha=1,2.
    \label{eq:Fluid_traction_to_solid}
\end{eqnarray}
with $n_j^{\textrm{F}+}=-n_j^{\textrm{F}-}=\delta_{3j}$ are the outward normal to the fluid domain at the top and bottom surfaces of the fracture.
The fluid shear stress at the walls $\sigma_{\alpha 3}\left(x_3=\pm \frac{w}{2}\right)$ in the lubrication approximation is given by  
\begin{equation}
    \sigma_{\alpha 3}\left(x_3=\pm \frac{w}{2}\right) =
    \pm \frac{w}{2} \left(\frac{\partial p}{\partial x_\alpha} -\rho^{\textrm{F}} g_\alpha\right) 
    ,\quad \alpha=1,2.
\end{equation}
We thus see that as a result of the fluid shear stress at the fracture wall, the fracture is not "self-equilibrated" as 
\begin{eqnarray}
    T_i^{\textrm{F}+}+T_i^{\textrm{F}-} = (\sigma_{ij}^+ - \sigma_{ij}^-)^{\textrm{F}} n_j^{\textrm{F}+} =
    w  
    \left(\frac{\partial p}{\partial x_\alpha} -\rho^{\textrm{F}} g_\alpha\right) \delta_{i \alpha} \ne 0,\quad \alpha=1,2, \\
    \frac{1}{2}\left(T_i^{\textrm{F}+} - T_i^{\textrm{F}-}\right) = 
     \frac{1}{2}(\sigma_{ij}^+ + \sigma_{ij}^-)^{\textrm{F}} n_j^{\textrm{F}+}= - p \delta_{i3},\quad i=1,2,3.
\end{eqnarray}
In a pressure-driven flow, the fluid thus transfers shear stress to the solid in a particular manner. Its impact on hydraulic fracture growth has been thoroughly investigated recently by \cite{WrPi17}. It is typically very small compared to the net pressure, given by the fluid pressure minus the initial compressive stress. This can be grasped by looking at its impact on solid deformation. We do so here for simplicity in a plane-strain two-dimensional configuration ($\mathbf{e}_1,\, \mathbf{e}_3$) for a planar fracture along $\mathbf{e}_1$. 
Recalling that $T_i^\textrm{S} = -T_i^\textrm{F}$ (and $n_j^{\textrm{F}+}=-n_j^{\textrm{S}+}=n_j^{\textrm{S}-}$), the  quasi-static balance of momentum for an elastic solid reduces to the following traction boundary integral equation for such a general loading (see for example \cite{PiMi13} or \cite{Mogi14} among many others):
\begin{eqnarray}
                  T_1^{\textrm{S}+}-T_1^{o+}=\frac{1-2\nu}{4(1-\nu)} 
         \frac{E}{4(1-\nu^2)} \mathcal{H}\left( \Delta u_1 \right), \\
         T_3^{\textrm{S}+}-T_3^{o+}=-\frac{1-2\nu}{4(1-\nu)} \mathcal{H}\left( T_1^{\textrm{F}+}+T_1^{\textrm{F}-} \right)
         +\frac{E}{4(1-\nu^2)} \mathcal{H}\left( \Delta u_3 \right) \label{eq:2d_el_2},        
\end{eqnarray}
with $\Delta u_i = u_i^+ - u_i^-$ the fracture displacement discontinuity and  $\mathcal{H}(f)$ the Hilbert transform defined on the planar crack as
\begin{equation}
    \mathcal{H} \left (f \right) (x_1) = \frac{1}{\pi} \int_\Gamma  \frac{f(\xi)}{x_1-\xi}  \textrm{d}\xi.
\end{equation}

In the thin-film lubrication limit, the fluid shear stress transmitted to the solid in the case of a pressure driven flow scales as  $ T_1^{\textrm{F}+}+T_1^{\textrm{F}-}\propto \tau  = W P /L $, where $W$ and $L$ are the characteristic width and length of the fracture as previously introduced. This directly stems from the fluid momentum balance for pressure-driven thin film flow and is independent of the fluid rheology. We thus see that the characteristic fluid shear stress is much smaller than the characteristic net pressure:
\begin{equation}
    \tau = \epsilon P 
\end{equation}
where $\epsilon = W/L$ is the characteristic aspect ratio of the fracture. 
The characteristic scales of the different terms of the elastic equation along $e_3$ \eqref{eq:2d_el_2} are thus respectively $P$ (for the net loading $T_3^{\textrm{S}+}-T_3^{o+}$),  $\epsilon P$ (for the effect of the fluid shear stress $T_1^{\textrm{F}+}+T_1^{\textrm{F}-}$), and $E^\prime W/L = E^\prime \epsilon $ (for the traction induced by the fracture opening $\Delta u_3$). As elasticity is always of first order, we recover the classical linear elastic relation between characteristic net pressure $P$ and characteristic strain associated with the fracture $\epsilon= W/L$:
\begin{equation}
    P = E^\prime W/L = E^\prime  \epsilon,
    \label{eq:P_characteristic}
\end{equation}.


If one accounts for the fluid shear stress, in the absence of shear due to far-field stresses, the elasticity equation along $e_1$ provides the following scale for $\Delta u_1$: $\delta = \tau L/ E^\prime = \epsilon^2 L $. In other words, the shear slip is of order $\mathcal{O}(\epsilon)$ compared to the fracture width. As a result, the part of the infinitesimal power associated with shear $T_\alpha \dot{u}_\alpha $ scales as $\tau \dot{\delta}= \epsilon P \dot{\widehat{(\epsilon^2 L)}} $ which is of order $\mathcal{O}(\epsilon^2)$ compared to the part related to the opening mode $P \dot{W} = P \dot{\widehat{(\epsilon L)}} $.
The fluid shear stress can thus be neglected as its magnitude is of order $\mathcal{O}(\epsilon)$ of the net pressure. Similarly,  the horizontal contribution to the power of external traction to the solid $T_\alpha \dot{u}_\alpha $ is negligible. In doing so, any associated shear slip is also negligible. This, however, does not preclude shear slip induced by fracture re-orientation due to pre-existing or transmitted shear stress $T_\alpha^o$ from the far-field (for example stress changes due to loading on the solid boundary $\partial \Omega_{}^{\textrm{S}}$). 

By neglecting the fluid shear stress, the fluid symmetrically loads the solid fracture surface. 
The fracture is self-equilibrated: $T_i^{\textrm{F}+}+T_i^{\textrm{F}-}=\mathcal{O}(\epsilon)\approx 0$. Finally, it is worth pointing out that this conclusion is not restricted to plane elasticity. The same arguments hold in 3D for a smoothly curved fracture although the details of the elastic operators are more lengthy. It is also not restricted to a particular fluid rheology and thus equally applies to a Newtonian or complex fluid.

\section{Energy balance for a linear hydraulic fracture at depth}
The global power balance for the linear hydraulic fracture model combines the energy equations for the solid \eqref{eq:Solid_balance_final} and the fluid under the lubrication approximation \eqref{eq:Fluid_P_lubrication}, using the interface condition \eqref{eq:F_S_interface_power}. 
In order to obtain the final expression, we further reduce the volumetric integrals appearing in the fluid volume balance to a surface integral over the middle surface of the fluid domain which coincides with the fracture middle surface $\Sigma$:
\begin{equation}
    \int_{\Omega^\textrm{F}} (\cdot) \textrm{d}V =  \int_{\Sigma} \int_{-w/2}^{w/2} (\cdot) \textrm{d}x_3 \textrm{d}S,
\end{equation}
where we recall that $w=\Delta u_3^S$ is the local fracture width.   In doing so, we introduce the horizontal mean fluid velocity
\begin{equation}
v_\alpha=\frac{1}{w}\int_{\,-w/2}^{\,w/2}\dot{u}^F_\alpha\,\textrm{d}x_3,\quad \alpha=1,2. \label{eq:Average_u}
\end{equation}
As a result, the effect of the fluid body forces can be written as:
\begin{equation}
     \int_{\Omega^\textrm{F}} \rho^{\textrm{F}} g_\alpha \dot{u}^{\textrm{F}}_\alpha \textrm{d}V = \int_{\Sigma} \rho^{\textrm{F}} g_\alpha w v_\alpha  \textrm{d}S, \quad \alpha=1,2.
\end{equation}
Integrating by part with respect to $x_3$ and using the fluid balance of momentum in the lubrication limit, the viscous dissipation becomes (see~\ref{app:Lubrication} for details):
\begin{equation}
    \dispdot{D}_{\textrm{dev}-0}^\textrm{F} = 
    \int_{\partial\Omega^{\textrm{S}\leftrightarrow \textrm{F}}} T^\textrm{F}_\alpha 
    \dot{u}^\textrm{F}_\alpha\,\textrm{d}S 
    + 12 \mu \int_{\Sigma} \frac{v_\alpha v_\alpha }{w} \textrm{d}S ,\quad \alpha=1,2.
\end{equation}
where the solution for pressure-driven lubrication flow has been used. The first integral in the expression $ \dispdot{\textrm{D}}_{\textrm{dev}-0}^\textrm{F} $ balances the horizontal component of the external power in the fluid power balance which thus reads
\begin{equation}
  Q_{in}\,p_{in}+  \int_{\partial\Omega^{\textrm{S}\leftrightarrow \textrm{F}}}{T_3^{F}\,\dot{u}^{F}_3\,\textrm{d}S}+ \int_{\Sigma} w \rho^F g_\alpha v_\alpha \textrm{d}S =  12 \mu \int_{\Sigma} \frac{v_\alpha  v_\alpha }{w} \textrm{d}S
  ,\quad \alpha=1,2.
\end{equation}

Neglecting the external power transmitted to the solid via the fluid shear stress as $T^S_\alpha \dot{u}^S_\alpha = \mathcal{O}(\epsilon^2)$ as previously discussed, the fracture is symmetrically loaded. The power balance of the solid reduces to:
\begin{equation}
\int_{\partial\Omega^{\textrm{S}\leftrightarrow \textrm{F}}}\dot{u}^\textrm{S}_3\,T_3^\textrm{S}\,\mathrm{d}S=
\int_{\Sigma}\Delta \dot{u}^\textrm{S}_i\,T_i^{o+}\,\mathrm{d}S +
\int_{\Sigma}\frac{\partial}{\partial t}\left[\frac{1}{2} \Delta u_i^\textrm{S}\,\left(T^\textrm{S+}_i-T_i^\textrm{o+}\right)\right]\mathrm{d}S+\int_{\Gamma}G_\textrm{c}\,v_i\,n_i\,\mathrm{d}\gamma,
\end{equation} 
with the solid displacement discontinuity $\Delta u_i=u_i^{\textrm{S}+}-u_i^{\textrm{S}-}$. At the solid-fluid interface: $T_3^\textrm{F} \dot{u}_3^\textrm{F} +\dot{u}^\textrm{S}_3\, T_3^\textrm{S}= T_3^\textrm{F} v_L \delta_{i3} $, such that the sum of the normal component of the external work exchange at the fluid-solid interfaces - expressing the integral over the fracture mid-plane - reduces to:
\begin{equation}
    \int_{\partial\Omega^{\textrm{S}\leftrightarrow \textrm{F}}}\dot{u}^\textrm{S}_3\,T_3^\textrm{S}\,\mathrm{d}S + \int_{\partial\Omega^{\textrm{S}\leftrightarrow \textrm{F}}}{T_3^{F}\,\dot{u}^{F}_3\,\textrm{d}S}
    = \int_\Sigma T_3^\textrm{F+}v_L^+ +   
    T_3^\textrm{F-} v_L^-\textrm{d}S 
    = - \int_\Sigma 2 p v_L \textrm{d}S, 
\end{equation}
on the account that the fluid leaks off symmetrically from the top and bottom surfaces $v_L^+= - v_L^-= v_L $.

Writing  $\Delta u_i=u_i^{\textrm{S}+}-u_i^{\textrm{S}-}= \Delta u_\alpha \delta_{i\alpha} + w \delta_{i3} $ to highlight the fracture aperture $w=\Delta u_3$, the sum of the fluid and solid power balances can be written as:
 \begin{eqnarray}
    \underbrace{Q_{in} p_{in}}_{\textrm{Injection}} - \underbrace{\int_\Sigma 2 p v_L \textrm{d}S}_{\textrm{fluid leak-off}} 
  + \underbrace{\int_{\Sigma} \rho^{\textrm{F}} g_\alpha w v_\alpha  \textrm{d}S}_{\textrm{fluid gravity}} \quad 
    = \\
    \underbrace{\int_{\Sigma} \dot{w} \sigma^o +  \Delta{\dot{u}}_\alpha  \tau^o_\alpha \textrm{d} S }_{\textrm{power spent against in-situ stress}}
    + \underbrace{\int_{\Sigma} \frac{\partial}{\partial t} \left(\frac{1}{2} \left( w (p - \sigma^o ) - \Delta{{u}}_\alpha  \tau^o_\alpha \right)  \right) \textrm{d} S}_{\textrm{rate of elastic energy}}+\
    \underbrace{12 \mu \int_{\Sigma} \frac{v_\alpha  v_\alpha }{w} \textrm{d}S}_{\textrm{viscous dissipation}} +
    \underbrace{\int_{\Gamma } G_c v_i n_i \textrm{d}\gamma }_{\textrm{fracture dissipation}},
    \label{eq:final_PB_1}
\end{eqnarray}
where we have expressed the effect of the 
initial compressive stress field on the fracture surface as $T_i^{o+}=\sigma^o \delta_{i3}+\tau^o_\alpha \delta_{i\alpha}$ ($\sigma^o,\,\tau^o_\alpha >0$ in compression). 

Hydraulic fractures are mode I fractures. They always re-orient to propagate normal to the minimum compressive stress as observed experimentally \citep{HuWi57,BuLe17}. Ultimately, in a uniform in-situ stress field, the preferred propagation plane corresponds to the one with zero shear stress $ \tau^o_\alpha=0$, and the normal stress $\sigma^o$ corresponds to the minimum principal stress. As a result, for a planar hydraulic fracture at depth propagating in the plane perpendicular to the in-situ minimum stress, the term in $ \Delta{u}_\alpha  \tau^o_\alpha$ disappears from the final power balance \eqref{eq:final_PB_1}. 

\subsection{Accounting for fluid volume conservation}
It is important to include fluid mass conservation in the ultimate power balance. For an incompressible fracturing fluid, the width integration of the fluid mass conservation is given by \cite{Deto16}: 
\begin{equation}
    \dot{w}
    + \frac{\partial w v_\alpha}{\partial x_\alpha} + 2 v_L = \delta(x_\alpha) Q_{in},
\end{equation}
assuming that the size of the wellbore from which the fluid is injected is negligible compared to the fracture size as done when deriving the power balance of the fluid. Multiplying by the in-situ stress normal to the fracture plane $\sigma^o$, integrating over the whole fracture surface $\Sigma$, we obtain the following expression for the work performed against the in-situ stress to open the fracture:
\begin{equation}
\int_{\Sigma}\dot{w}\,\sigma^\textrm{o}\,\mathrm{d}S=Q_{in}\,\sigma^\textrm{o}-2\int_{\Sigma}\sigma^{\textrm{o}}\,v_L\,\textrm{d}S+\int_{\Sigma}\frac{\partial \sigma^{\textrm{o}}}{\partial x_\alpha} w v_\alpha\,\mathrm{d}S, \quad \alpha=1,2,\label{eq:rate_conf_stress}
\end{equation}
where we have used the boundary conditions of zero fluid flux $w v_\alpha$ at the fracture front.

This allows to re-write the global power balance in terms of the net pressure $p_{net}=p-\sigma^o$: the fluid pressure in excess of the minimum in-situ confining stress. Restricting to the case of a planar fracture oriented in the maximum principal stress direction such that $\tau^o_\alpha=0$, we obtain the final power balance for a planar hydraulic fracture at depth:
 \begin{eqnarray}
    \underbrace{Q_{in} (p_{in}-\sigma^o(0))}_{\textrm{Injection}}  
  + \underbrace{\int_{\Sigma} w v_\alpha \left(\rho^{\textrm{F}} g_\alpha  -\frac{\partial \sigma^{\textrm{o}}}{\partial x_\alpha} \right) \textrm{d}S}_{\textrm{buoyancy}}  
    = \nonumber \\
    \underbrace{\int_{\Sigma} \frac{\partial}{\partial t} \left(\frac{1}{2}  w p_{net} \right) \textrm{d} S}_{\textrm{rate of elastic energy}}+
     \underbrace{\int_\Sigma 2 p_{net} v_L \textrm{d}S}_{\textrm{fluid leak-off}}+
    \underbrace{12 \mu \int_{\Sigma} \frac{v_\alpha  v_\alpha }{w} \textrm{d}S}_{\textrm{viscous dissipation}} +
    \underbrace{\int_{\Gamma } G_c v_i n_i \textrm{d}\gamma }_{\textrm{fracture dissipation}}.
    \label{eq:final_PB_2}
\end{eqnarray}
The power input coming from fluid injection and buoyancy are partly stored in elastic deformation and dissipated via fluid leak-off in the surrounding medium, viscous fluid flow and the creation of new fracture surfaces. The corresponding energy balance can be chiefly obtained by simple time-integration of the power balance from the start of the injection. 

\subsection{Case of a fluid lag}
In the case of low confining stress / highly viscous fluid, the fracturing fluid front may lag behind the fracture front (see \cite{GaDe00,Gara06c,LeDe07,BuGo13,DePe14}). When the medium is impermeable, fluid vapour (at the cavitation pressure) fills this (evolving) tip cavity, while in porous rock reservoirs fluid fills such cavity \citep{DeGa03,KaGa19}. The presence of a fluid lag can be easily accommodated by splitting the integrals over the fluid-filled and lag parts of the fracture. An additional term associated with the fluid front moving boundary $\int_{\Gamma_f} \sigma^o w v^f_\alpha n_\alpha \textrm{d}S $ appears in the expression of the work performed against in the in-situ stress \eqref{eq:rate_conf_stress}. 
We refer to \cite{LeDe07} for more details. Theoretical and experimental works have demonstrated that the fluid and fracture fronts coalesce over a time-scale $t_{om}=12 E^{\prime2} \mu / (\sigma^o)^3 $. Such a time scale is usually small for a realistic value of $\sigma^o$ at depth, and the lag can thus be neglected in most practical cases.

\subsection{Carter's leak-off model}
We have so far not expressed explicitly the fluid leak-off velocity $v_L$. The simplest model assumes one-dimensional Darcy's flow perpendicular to the fracture plane symmetrically from the top and bottom surfaces. If in addition, the reservoir is normally pressurized such that the initial effective stress is significantly larger than the propagating net pressure (the fluid pressure minus the compressive stress), the leak-off rate is nearly independent of the fluid pressure inside the crack. Under those assumptions, Carter's leak-off model applies \citep{HoFa57,LeZh18,KaGa19}. It states that the leak-off velocity $v_L$ is inversely proportional to the square root of the effective time of exposure $(t-t_\textrm{o})$, where $t_\textrm{o}$ is the time at which the fracture opens at a given location $(x_1,x_2,x_3)$. Under those approximations, the leak-off velocity $v_L$ is expressed as 
\begin{equation}
v_L=\frac{C_L}{\sqrt{t-t_\textrm{o}(x_i)}},\label{eq:Carter_leakoff}
\end{equation}
where $C_L$ is the Carter's leak-off coefficient which is a function of the rock permeability, storage, fluid viscosity, and the reservoir effective stress (see \cite{LeZh18} for further details). Carter's leak-off model is widely used and is part of the set of governing equations of linear hydraulic fracture mechanics \citep{Deto16}. 

\section{Scaling \& Competition between the different powers at play}

In its form \eqref{eq:final_PB_2},  the power balance has been expressed using the Newtonian lubrication solution of the fluid balance of momentum (for the fluid viscous dissipation), as well as the fluid mass balance. When combined with the solid quasi-static balance of momentum and the expression of the leak-off velocity \eqref{eq:Carter_leakoff}, the power balance provides the complete description of hydraulic fracture propagation. We can then recover the scaling and understanding of the different hydraulic fracture propagation regimes in an intuitive way based on the different powers.

First, we recall that the solid balance of momentum provides the relation between the characteristic net pressure $P$, characteristic fracture width $W$ and characteristic fracture size $L$ (see Eq.~\eqref{eq:P_characteristic}):
\begin{equation}
    P= \epsilon E^\prime,\, \epsilon = W/L.
\end{equation}
In the absence of a fluid lag, the fluid velocity $v_\alpha$ scales similarly to the fracture velocity $v_i n_i$ as $L/t$. We can thus write the characteristic scale of the different power terms of the balance \eqref{eq:final_PB_2} as follows.
\begin{itemize}
    \item The energy input scales as $P\, Q$, where $Q$ is the characteristic inflow rate.
    \item The effect of the buoyancy contrast ($\Delta \gamma = \rho^{\textrm{F}} g_\alpha  -\dfrac{\partial \sigma^{\textrm{o}}}{\partial x_\alpha} $) scales  as $  \text{Area}  \times W L \Delta \gamma /t $, where $\text{Area}$ denotes the characteristic area of the fracture.
    \item The rate of stored elastic energy scales as $\text{Area}  \times   W P / t$. 
    \item The leak-off power scales as $\text{Area}  \times C^\prime \sqrt{t} L  $, with $C^\prime = 2 C_L$ (Carter's leak-off model).
    \item The viscous dissipation scales as $\text{Area}\times \mu^\prime L^2/(W t^2) $ with $\mu^\prime =12 \mu $.
    \item The energy dissipated in the creation of new fracture surfaces scales as $\text{Perimeter}\times K_{Ic}^2 L / (E^\prime t) $, where we have used Irwin's relation for a strictly tensile mode I fracture relating the critical fracture energy to the mode I fracture toughness: $G_c=K_{Ic}^2 / E^\prime$.
\end{itemize}

\subsection{Propagation regimes and scaling for radial hydraulic fracture}
\label{subsec:Prop-Regimes}
For a radial geometry ($\text{Area}=L^2$, $\text{Perimeter}=L$), we first focus on the case with negligible buoyancy contrast ($\Delta \gamma \sim 0$).  
Dividing the different power terms by the characteristic power input, 
we recover the classical dimensionless groups for a radial hydraulic fracture \citep{SaDe02,Deto04,Ma04,Deto16}:
\begin{itemize}
    \item the ratio of stored elastic energy over the power input: $\mathcal{G}_v = \dfrac{\epsilon L^3}{Q t} $, which strictly corresponds to the ratio of the fracture volume over the characteristic injected volume.
    \item the ratio of the leak-off dissipation over the power input: $\mathcal{G}_c = \dfrac{C^\prime L^2 }{Q \sqrt{t}} $, which similarly corresponds to the ratio of the leaked-off volume over the characteristic injected volume.
    \item the  viscous dissipation over the power input 
    $\mathcal{G}_m=\dfrac{\mu^\prime L^3}{\epsilon^2  Q t^3}$
    \item and the fracture dissipation over the power input
    $\mathcal{G}_G=\dfrac{K_{Ic}^2 L^2}{\epsilon E^{\prime 2} Q t}$.
\end{itemize}
These four dimensionless coefficients define the propagation diagram for a radial hydraulic fracture, with four distinct limiting regimes: storage-viscosity, storage-toughness, leak-off-viscosity and leak-off-toughness.
If most of the injected fluid is stored in elastic deformation, and viscous dissipation dominates over the creation of fracture surfaces, we set $\mathcal{G}_v=\mathcal{G}_m=1$ and recover the viscosity-storage scaling (subscript $m$):
\begin{eqnarray}
    \epsilon_m = \left(\frac{\mu^\prime}{E^\prime t}\right)^{1/3},\qquad L_m= \frac{E^{\prime 1/9} Q^{1/3} t^{4/9}}{\mu^{\prime 1/9}}, \\
\mathcal{G}_c \equiv \mathcal{C}_m(t)= C^\prime \frac{E^{\prime 2/9} t^{7/18} }{Q^{1/3} \mu^{\prime 2/9} },\qquad  \mathcal{G}_G \equiv \mathcal{K}^2_m(t)=\frac{K_{Ic}^2 t^{2/9}}{E^{\prime 13/9} Q^{1/3} \mu^{\prime 5/9}}, 
    \label{eq:M-scaling}
\end{eqnarray}
where $\mathcal{C}_m$ is the dimensionless leak-off coefficient first defined in \cite{Mady03}, and the dimensionless fracture energy is equal to the square of the dimensionless fracture toughness $\mathcal{K}_m$ first defined in \cite{SaDe02}. We observe that the fracture energy dissipation increases with time (in relation to the increase of the perimeter of such radial fracture). Similarly, the effect of fluid leak-off increases over time. The times at which the dimensionless toughness and dimensionless leak-off become of order one provide two time scales:
\begin{eqnarray}
    \mathcal{C}_m = \left( t/t_{m\tilde{m}} \right)^{7/18}, \qquad  t_{m\tilde{m}} = \frac{\mu^{\prime 4/7}Q^{6/7}}{E^{\prime 4/7}C^{\prime 18/7 }},
    \label{eq:Dimless-Leak-Off-tmmt}
    \\
    \mathcal{K}_m = \left( t/t_{mk} \right)^{1/9}, \qquad t_{mk}=\frac{\mu^{\prime 5/2}Q^{3/2} E^{\prime 13/2}}{K_{Ic}^9}.
    \label{eq:Dimless-Toughness-tmk}
\end{eqnarray}
The viscosity-storage regime corresponds to the early stage of propagation, where both the effect of fluid leak-off and fracture energy are small compared to storage and fluid viscous dissipation: $\mathcal{C}_m\ll 1,\,\mathcal{K}_m\ll 1$.
At large times, the hydraulic fracture evolves toward the leak-off / toughness-dominated regime where both fluid storage and viscous flow dissipation in the crack become negligible. If $t_{mk}\ll t_{m\tilde{m}}$, the hydraulic fracture will first evolve toward the storage toughness-dominated regime (where leak-off and viscous dissipation are negligible), while if $t_{mk}\gg t_{m\tilde{m}}$, it will pass via the leak-off viscosity regime. By alternately setting either the storage or leak-off and the viscous versus the fracture energy as the dominant terms of the energy balance, one can recover the different scalings corresponding to the different propagation regimes of a hydraulic fracture. We refer to \cite{Deto16} for a detailed discussion and the expressions of the different scales in the different limiting regimes. 

It is important to note that the coefficients  (toughness, leak-off) obtained by scaling arguments from the strong form of the linear hydraulic fracture mechanics model are indeed quantifying the order of magnitude of the different power terms - which is more clearly seen from the energy approach taken here. 

Another point worth noting is that in previous contributions an equivalent toughness $K^\prime = \sqrt{32/\pi} K_{Ic}$ was used in the definition of the different propagation scalings. Such a pre-factor ($\sqrt{32/\pi}$) comes from the LEFM near-tip asymptotic in plane-strain for the crack opening displacement. However, the use of $K^\prime$ instead of $K_{Ic}$ grossly distorts the transition between the viscosity- and the toughness-dominated regime which does not occur at a dimensionless toughness of $1$ if $K^\prime$ is used instead of $K_{Ic}$. This can be directly related to the fact that the pre-factor $\sqrt{32/\pi}$ does not appear in the power balance. On the other hand, the use of the equivalent viscosity $\mu^\prime=12 \mu $ and leak-off coefficient $C^\prime = 2 C_l$ is warranted as these pre-factors appear directly in the power balance \eqref{eq:final_PB_2}, and therefore impact the order of magnitude of the corresponding dissipations.

\paragraph{Buoyancy}

If we include the  buoyancy term $\Delta \gamma \ne 0$, the ratio of the characteristic buoyant power over the characteristic rate of stored elastic energy for a radial geometry reads
\begin{equation}
    \mathcal{G}_b = \frac{\Delta\gamma L}{P},
\end{equation}
and one recovers for the storage viscosity-dominated regime the dimensionless 
buoyancy defined in \cite{MoLe22}:
\begin{equation}
    \mathcal{G}_b \equiv  \mathcal{B}_m = \frac{\Delta \gamma Q^{1/3}t^{7/9}}{E^{\prime 5/9} \mu^{\prime 4/9}},
    \label{eq:Dimless-Buoyancy}
\end{equation}
which governs the onset of the buoyant regime in the viscosity-dominated radial regime. We refer to \cite{MoLe22,MoLe23} and references therein for a detailed discussion of buoyant fracture growth.

\subsection{Plane-strain geometry}  
For a plane-strain geometry (setting $\text{Area}=L$, $\text{Perimeter}=1$), one can similarly recover the dimensionless coefficients, limiting regimes and corresponding scaling previously obtained in \cite{Adac01,AdDe02,AdDe08,GaDe05,Gara06b,Gara06c}. The main difference compared to the radial/3D case is that because the "perimeter" of the fracture reduces to a point in plane strain, the dimensionless fracture energy is constant and does not evolve with time for the plane-strain geometry. We refer to \cite{Deto04,Gara06b,HuGa10} for details.


\subsection{Elongated / blade-like hydraulic fracture}

Another interesting geometry corresponds to a blade-like hydraulic fracture propagating with a constant breadth $H$ much smaller than its other propagating dimension $L$. Such a geometry - often referred to as PKN in the petroleum literature - corresponds to a 3D hydraulic fracture with its propagation contained within one rock layer of thickness $H$. For such a geometry, the propagating fracture front is of scale $H$, such that the characteristic fracture area and perimeter are: $\text{Area}=LH$,  $\text{Perimeter}=H$. Because the fracture is contained, the characteristic elastic strain $\epsilon$ must be defined with the fracture breadth $H$: $\epsilon = W/H$. This is in agreement with the fact that the elastic relation between net pressure and width is local for sufficiently elongated fracture $L\gg H$ (see \cite{AdPe08,DoPe15,Dont21} for discussion). With these definitions for the characteristic fracture area, perimeter and elastic strain, the power balance provides the  corresponding dimensionless power ratios for such a geometry:
\begin{eqnarray}
    \mathcal{G}_v = \frac{\epsilon L H^2}{Q t}, \qquad \mathcal{G}_c=\frac{C^\prime H L }{ Q \sqrt{t}}, \\
    \mathcal{G}_m =\frac{\mu^\prime L^3}{ \epsilon^2 E^\prime Q t^2}, \qquad \mathcal{G}_k = \frac{K_{Ic}^2 H L }{\epsilon E^{\prime 2} Q t}. 
\end{eqnarray}
It allows in a strictly similar manner as for the radial geometry to recover the propagation regimes for such a contained hydraulic fracture - see \cite{Dont21} for details from the strong form. 

Notably, the blade-like hydraulic fracture initially propagates in a regime dominated by elastic storage ($\mathcal{G}_v=1$) and the energy dissipated in the creation of new fractures  ($\mathcal{G}_G=1$) -  (subscript $k$ for toughness dominated)
\begin{eqnarray}
    \epsilon_k = \frac{K_{Ic}}{E^\prime \sqrt{H}} ,\qquad L_k = \frac{E^\prime Q t}{H^{3/2} K_{Ic}}, \\ 
    \mathcal{G}_c \equiv \mathcal{C}_k=\frac{C^\prime E^\prime \sqrt{t}}{K_{Ic}\sqrt{H}},\qquad \mathcal{G}_m=\mathcal{M}_k= \frac{\mu^\prime E^{\prime 4 } Q^2 t }{H^{7/2} K_{Ic}^5},
\end{eqnarray}
where we observe that the dimensionless energy dissipated in viscous fluid flow increases as the fracture grows and its aspect ratio $L/H$ increases. We also can note, that the characteristic pressure $\epsilon_k E^\prime $ and width $\epsilon_k H $ are constant in time in this early time storage-toughness regime. At late time, viscous dissipation thus dominates. In the absence of fluid-leak-off, we recover the classical PKN scaling first obtained in \cite{PeKe61,Nord72,Kemp90} (where the characteristic width and pressure increases with time). Similarly, when fluid leak-off becomes of order one, leak-off toughness and viscosity regimes can be obtained. We refer to \cite{Dont21}  for the corresponding expressions and approximate solutions for the contained fracture geometry.

\begin{figure}
\noindent \begin{centering}
\includegraphics[width=0.6\textwidth]{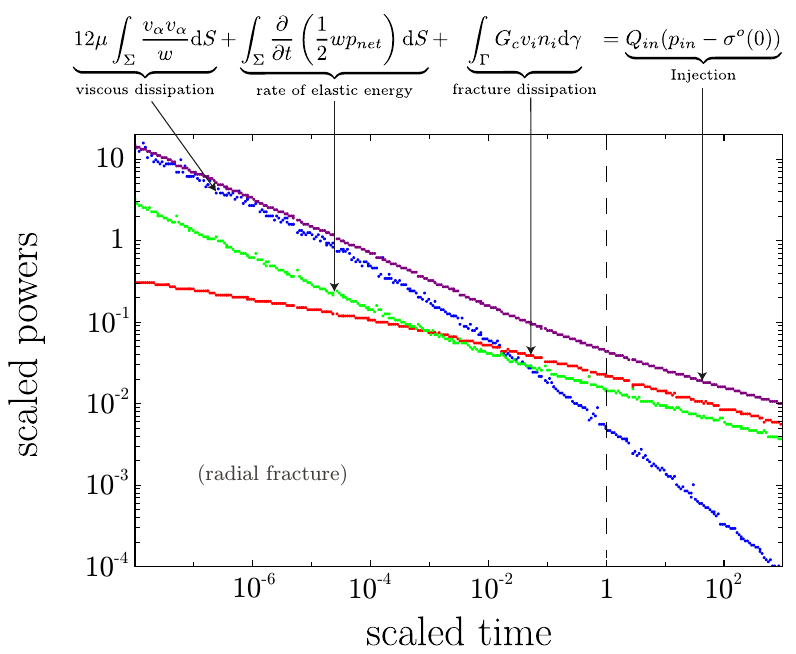}
\par\end{centering}
\caption{Scaled terms of the power balance as a function of the dimensionless time $t/t_{mk}$~\eqref{eq:Dimless-Toughness-tmk} for a radial hydraulic fracture transitioning from the early-time viscosity- to the late-time toughness-dominated regime.\label{fig:EB}}
\end{figure}
%
\section{Examples}

We provide hereafter several examples of hydraulic fracture propagation under different regimes highlighting the evolution of the different powers at play. All the examples were obtained using an open-source solver for the propagation of three-dimensional planar hydraulic fracture \citep{ZiLe20}. From the numerical results, we recomputed the different terms of the power balance~\eqref{eq:final_PB_2} numerically at each time step.

\subsection{Radial hydraulic fracture without fluid leak-off}

As a first example, we consider one of the most studied forms of hydraulic fractures, the radial or so-called penny-shaped fracture. Such fractures propagate from a point source in a homogeneous and impermeable medium. For simplicity, we exclude the work associated with overcoming a uniform and constant confining stress $\sigma_{\textrm{o}}$  in the power balance~\eqref{eq:final_PB_2} and write it using the fluid net pressure $p_{net}$. 

\subsubsection{Viscosity to toughness dominated growth \label{subsub_MK}}

In the absence of fluid leak-off, a radial hydraulic fracture driven by a continuous injection rate has been shown by~\cite{SaDe02} to transition from an early-time viscosity- to a late-time toughness-dominated regime. These two limiting regimes have self-similar solutions \citep{SaDe02,AbMu76}. The evolution of the solution is captured by a single dimensionless parameter, as discussed in Sec.~\ref{subsec:Prop-Regimes}. We use here the dimensionless toughness $\mathcal{K}_m$ as given in Eq.~\eqref{eq:Dimless-Toughness-tmk} or equivalently the characteristic transition time $t_{mk}$. Fig.~\ref{fig:EB} shows the time evolution of the different terms of the power balance. These terms are named directly in Fig.~\ref{fig:EB}, where all have been scaled by the characteristic input power $Q \sigma^o$.  We can clearly observe that the dominant energy dissipation mechanism at early time (small $t/t_{mk}$) is the viscous fluid flow (blue dots), while at later time (large $t/t_{mk}$) the energy spent to create new surfaces (red dots) become prominent. Interestingly, the transition time ($t=t_{mk}$)  does not mark the exact location of the cross-over between the viscous (blue) and surface creation (red) powers but rather the end of the transition. It corresponds to the time when the toughness-dominated regime is already well established.


\begin{figure}
\noindent \begin{centering}
\includegraphics[width=0.48\textwidth]{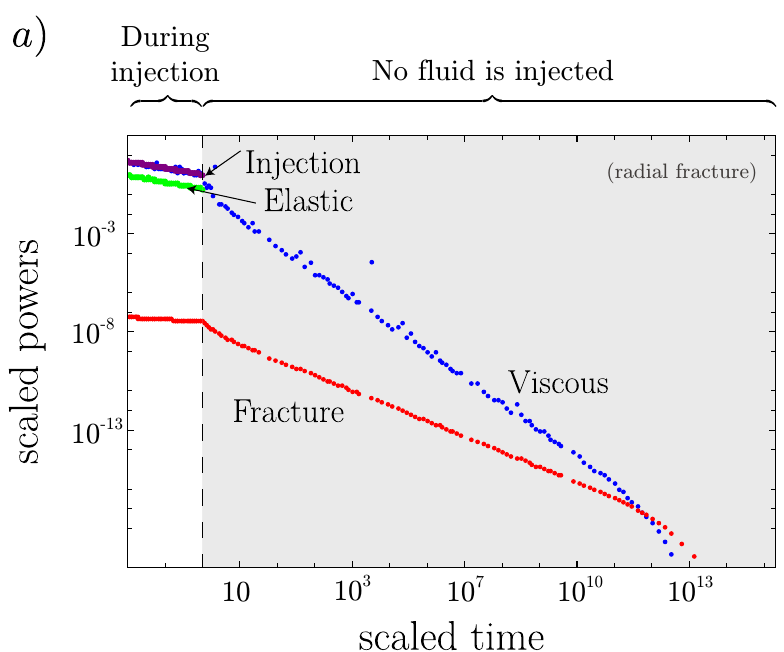}
\vspace{0.01\textwidth}
\includegraphics[width=0.48\textwidth]{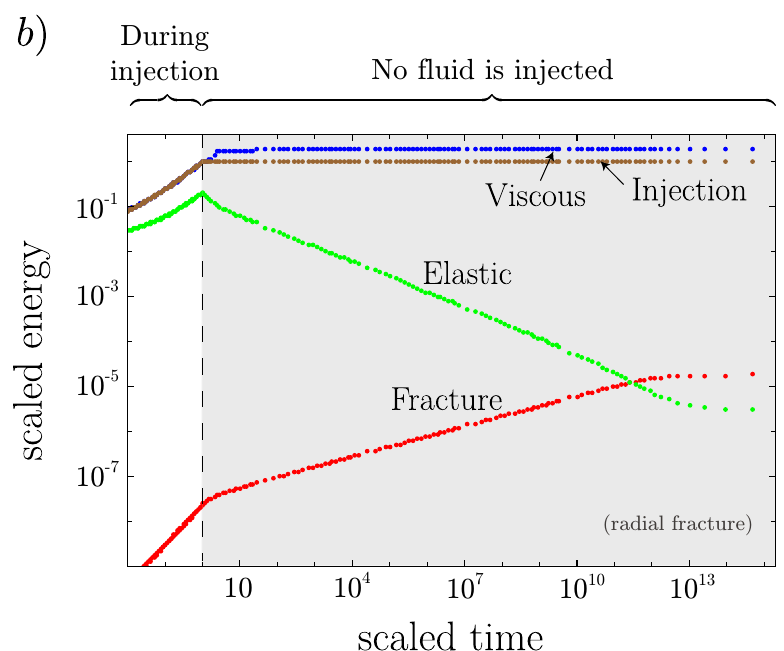}
\par\end{centering}
\caption{Scaled terms of the power balance as a function of the scaled time $t/t_s$, with $t_s$ the time when the injection stops, for a radial hydraulic fracture of finite injected volume. a) Power balance, b) Energy balance for the same case. Color code as in a).\label{fig:MPulse}}
\end{figure}

\subsubsection{Viscosity dominated pulse propagation}

The previous example considered that the injection continues at a constant rate without ever stopping. In any instance of a hydraulic fracture, may it be natural or anthropogenic, the fluid injection will necessarily stop after some time. \cite{MoLe21} have investigated in detail the case where the injection stops abruptly with a drop of the injection rate from $Q_{in}$ to $0$. 
When the injection stops while the hydraulic fracture propagates in the viscosity-dominated regime,  the hydraulic fracture can continue its propagation in a so-called pulse-like regime $\text{M}^{\left[V\right]}$. We highlight the source of this post-injection propagation in the example of Fig.~\ref{fig:MPulse}. Fig.~\ref{fig:MPulse}a) shows the power balance during injection, and the grey-shaded area  of  shows the dissipations after the injection has stopped. No more power enters the system after that time (as $Q = 0$). The elastic power is not visible after the injection has stopped in this log-log plot as it becomes negative. The excess of energy stored elastically during injection while the fracture propagates in the viscous regime is used to further drive the growth after shut-in. This is better grasped when looking at the evolution of the different energy terms (Fig.~\ref{fig:MPulse}b)). After the injection has stopped, one observes that the elastic energy decreases, showing a release of elastic energy. On the contrary, the energy spent to create new surfaces increases, consuming part of the released elastic energy in the system. The other part is consumed in viscous energy, and due to the different orders of magnitude of the cumulative energies, this can be clearly appreciated only from the plot of the different powers in Fig.~\ref{fig:MPulse}a). Of course, the fracture growth finally stops and the different powers all go to zero: the different energy terms stabilize at a final value as seen in Fig.~\ref{fig:MPulse}b). 

\begin{figure}
\noindent \begin{centering}
\includegraphics[width=0.75\textwidth]{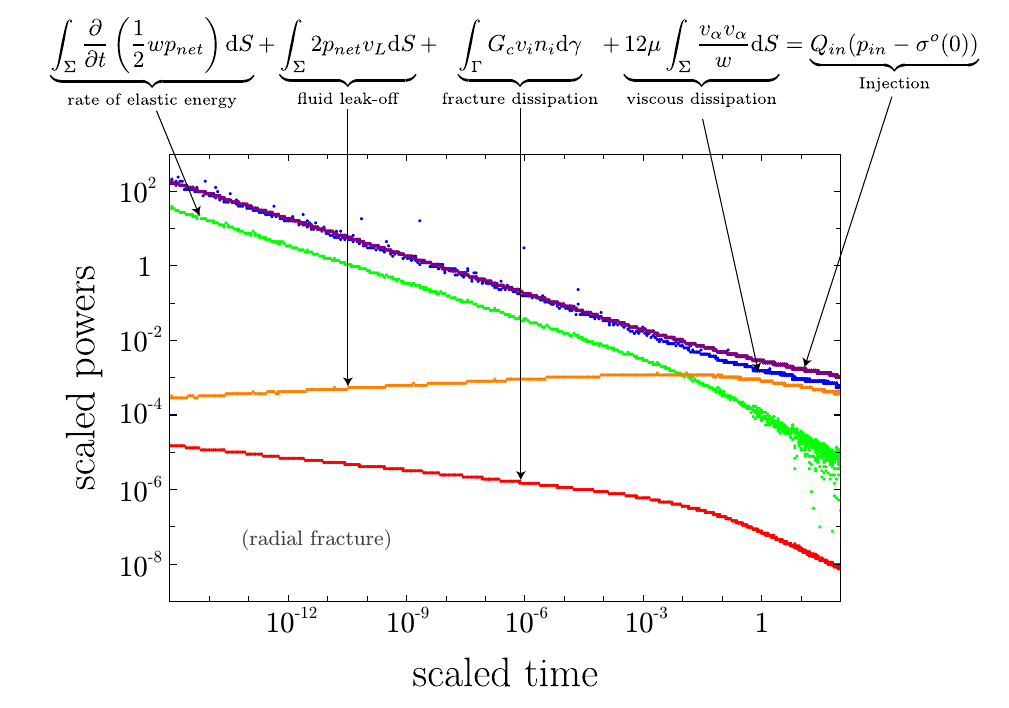}
\par\end{centering}
\caption{Example of a radial hydraulic fracture transitioning from the storage-viscosity ($M$)  to the leak-off-viscosity ($\tilde{M}$) regime. Scaled terms of the power balance as a function of dimensionless time $t/t_{m\widehat{m}}$~\eqref{eq:Dimless-Leak-Off-tmmt} for a radial hydraulic fracture with leak-off and a continuous injection.\label{fig:EB_leakOff}
}
\end{figure}

\subsection{Radial hydraulic fracture with leak-off}

Hydraulic fracture growth is substantially impacted by fluid leak-off. For a radial geometry, the propagation of the fracture depends on an additional dimensionless coefficient which is either a dimensional storage or dimensionless leak-off coefficient \citep{Mady03,AdDe08,HuGa10}.  Generally, the latter is chosen to describe the fracture (see  Eq.~\eqref{eq:Dimless-Leak-Off-tmmt}). The propagation path from the viscosity storage to the toughness-leak-off regime is summarized by a single trajectory parameter given by a combination of the two governing coefficients \citep{Deto16}. We display in Fig.~\ref{fig:EB_leakOff} the evolution of the different terms of the power balance \eqref{eq:final_PB_2} as a function of dimensionless time from a viscosity-storage (early time) to the viscosity-leak-off dominated regime (intermediate regime). The dimensionless toughness remains small throughout this simulation. Fig.~\ref{fig:EB_leakOff} highlights that the fluid leak-off gains importance with time and overcomes fluid viscous flow as the main energy dissipation mechanism. For such continuous injection, a late-time cross-over between the viscous and fracture energy dissipation similar to the one observed in Fig.~\ref{fig:EB} would ultimately appear if the simulation was pursued. As this phenomenon has already been demonstrated in subsection~\ref{subsub_MK}, we only highlight the nature of the leak-off dissipation here. 

\begin{figure}
\noindent
\includegraphics[width=\textwidth]{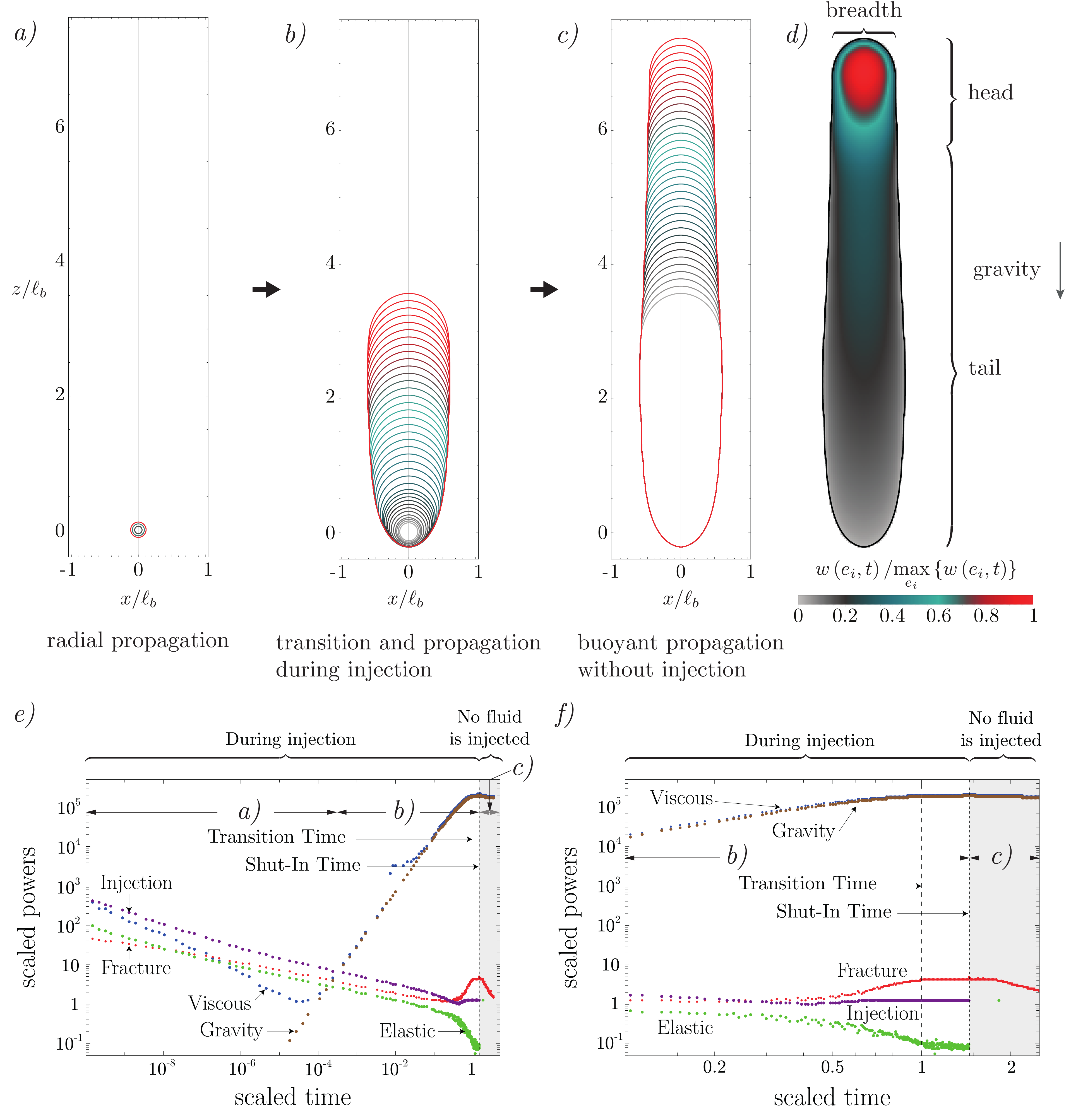}
\caption{Footprints and scaled terms of the power balance as a function of the scaled time $t/t_{k\widehat{k}}$ (with $t_{k\widehat{k}} = K_{Ic}^{8/3}/\left(E^\prime Q\varDelta\gamma^{5/3}\right)$) for a buoyancy-driven hydraulic fracture of finite injected volume. a) to c) are evolutions of the footprint from initially radial a) to finally buoyant without injection c). d) Shows the opening distribution of a buoyant fracture to highlight the distinction between fracture head and tail. e) Evolution of the different terms in the power balance (with notably  the one related to gravity). f) Zoom onto the phase of transition and propagation after the end of the injection of f), showing non-monotonic behaviours in various power terms.\label{fig:EB-Buoyancy}}
\end{figure}

\subsection{Buoyant 3D hydraulic fractures with a finite injection} 

The previous examples of radial hydraulic fractures had monotonic evolution of the different energy dissipation mechanisms. We now discuss a more complex hydraulic fracture geometry where buoyant effects become dominant and, as a result, elongate the fracture in the direction of gravity. Fig.~\ref{fig:EB-Buoyancy} shows the evolution of the fracture footprints (a)-c)) and of the dissipated powers (e) and f)) when considering buoyant forces. Fig.~\ref{fig:EB-Buoyancy}c) shows the entire history of the evolution of the different power terms from the initially radial propagation (see Fig.~\ref{fig:EB-Buoyancy}a)) up to the late-time, finite volume buoyant propagation (Fig.~\ref{fig:EB-Buoyancy}b) for the early time buoyant propagation - Fig.~\ref{fig:EB-Buoyancy}c) for the late-time buoyant propagation). The dissipation terms are shown as a function of the dimensionless transition time from radial to buoyant propagation $t_{k\widehat{k}}$ which can be derived from Eq.~\eqref{eq:Dimless-Buoyancy} and found in~\cite{MoLe22c}. At about $t/t_{k\widehat{k}} \approx 10^{-7}$ the transition from radial viscosity to radial toughness dominated can be observed as shown and described in Fig.~\ref{fig:EB}. The additional effect of the gravitational force (brown dots) then starts to act at about $t/t_{k\widehat{k}} \approx 5 \cdot 10^{-4}$. It is mainly consumed by fluid flow becoming predominantly in the buoyant direction. When the transition has ended, fracture propagation becomes uni-directional.
The fracture takes an elongated shape characterized by a constant breadth, a growing tail, and a moving "head" with fixed geometry \citep{MoLe22c} (see Fig.~\ref{fig:EB-Buoyancy}d)). As a consequence, the energy dissipation spent in the creation of new surfaces becomes constant. This is well observed in Fig.~\ref{fig:EB-Buoyancy}f) where the red line directly at the end of the transition ($t/t_{k\widehat{k}} \approx 1$) becomes constant. 
\cite{MoLe23} have further shown that after the end of injection, if the injected volume is above a critical threshold, these fractures continue to grow thanks to the energy input from the gravitational body force. We clearly observe this phenomenon in Fig.~\ref{fig:EB-Buoyancy}e). Even though the injection power (purple dots) goes to zero (the injection has stopped), gravitational power (brown dots) is still available to drive the fracture. The exact mechanism here is that gravity pushes fluid flow upwards and large portions of the fracture tail start to deplete. This depletion in the fracture tail releases elastic energy stored in the system, which in turn allows the fracture to continue its growth, similar to the post-injection propagation observed in Fig.~\ref{fig:MPulse}.

\subsection{Breakthrough of an initially contained hydraulic fracture}

\begin{figure}
\noindent \begin{centering}
\includegraphics[width=.8\textwidth]{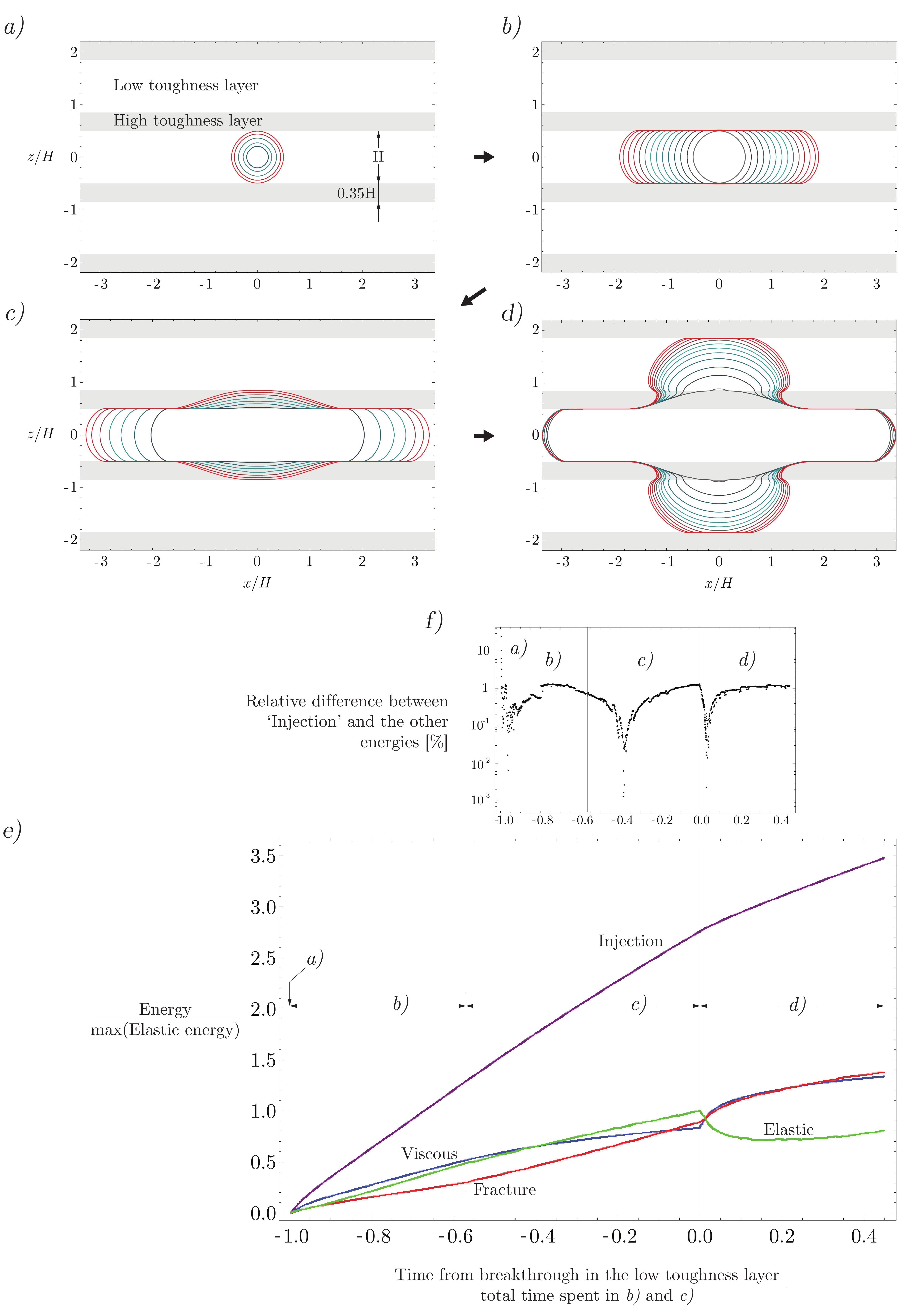}
\par\end{centering}
\caption{Interaction of a hydraulic fracture and layers of varying fracture toughness. The sub-figures a) to d) present the footprint evolution during different time intervals of the fracture propagation. Sub-figure e) present the evolution of the different terms of the energy balance. f) presents the relative difference between the injected energy and the sum of the other energies. \label{fig:EV}}
\end{figure}

The dominant energy dissipation mechanism can vary during hydraulic fracture propagation. 
This can be well observed in the presence of heterogeneities. 
We consider the simple case of an initially radial hydraulic fracture that encounters (symmetrically top and bottom) alternative layers  of higher then lower fracture toughness. The fracture footprint evolution resulting from such a numerical simulation is presented in Fig.~\ref{fig:EV}, and it is split into figures a) to d). The behaviour of the different terms of the energy balance reflects the hydraulic fracture interaction with such a heterogeneity. When the radial fracture reaches the high toughness layers the fracture is first contained in the lower toughness layer, thus propagating only in the lateral direction Fig.~\ref{fig:EV}b). During this  phase of propagation, the fracture energy (red curve) increases at a lower pace than the viscous one (blue curve). In fact, as soon as the fracture spreads laterally the fluid area increases, while the size of the propagating front remains constant. As the fracture gets longer, the pressure drop between the injection and the propagating front increases as well. The pressure increase at the injection point causes the fracture to eventually breakthrough in the high toughness region Fig.~\ref{fig:EV}c), explaining both the sudden rise in the power spent in creating new fracture surfaces (slope of the red curve) and the drop in both, viscous and elastic (green curve) powers.
When reaching the lower toughness layers, the hydraulic fracture accelerates due to the excess of elastic energy (Fig. \ref{fig:EV}d). Both the viscous and the fracture powers increase rapidly at the expense of the elastic energy stored in the rock. Finally, the fracture is again stopped by the next high-toughness layers such that the propagation takes place mainly in the low toughness layers. 

\section{Energy dissipation in practice}

\subsection{Additional dissipations: Wellbore flow, perforations and near-wellbore losses}
    
The power balance \eqref{eq:final_PB_2} was obtained assuming that the injection reduces to a point. In practice, fluid is delivered via a wellbore which can be several kilometers long while its diameter typically lies in a range $[8-30]$ centimeters. To initiate hydraulic fractures from a cased and cemented wellbore, shaped charges are detonated to create so-called perforations across the casing, cement reaching the rock formation. Hydraulic fracture initiation from such perforations can be quite complex but the hydraulic fracture re-orients and aligns with the preferred propagation direction dictated by the initial in-situ stresses over a distance equal to about three to four wellbore diameter (about 1 meter) (see \cite{BuLe17} for a more in-depth discussion).

Although turbulent flow may occur in the wellbore, the flow inside the fracture is laminar except at very early times (see \cite{Gara06,ZiLe17,LeZi19,GeGr24} for discussion). The fluid flow in the wellbore is well-modelled by pipe flow hydraulics. After the early time pressurization phase of the wellbore and the subsequent fracture initiation, quasi-steady-state flow conditions in the wellbore prevail under a constant injection rate $Q$ at the pump. We can thus write the following power balance for the fluid flow in the wellbore of total curvilinear length $L$ between the pump on the surface and the injection point at depth where fluid enters the hydraulic fracture:
    \begin{eqnarray} 
        \left(p_{\textrm{surf}} Q- p_{in} Q \right) 
        +\underbrace{\int_{0}^L \pi R^2(s) \rho^F g \sin \alpha(s) V(s) \text{ d}s}_{\textrm{gravitational power}} = 
        \underbrace{\int_0^{L} 2\pi R(s) \tau_w(s) V(s) \text{ d}s}_{\textrm{viscous friction}}  \nonumber\\
        +  \underbrace{\frac{K_p}{2} \rho^F V_{in}^2 Q}_{\textrm{perforations friction}} + \underbrace{K_{nwb} V_{in}^\beta Q}_{\textrm{near-wellbore friction}} + 
       \underbrace{ \int_0^L  V(s) \frac{\rho^F}{2} \frac{\partial \pi R(s)^2 V(s)^2}{\partial s} \text{ d}s    }_{\textrm{Kinetic losses}}
       \label{eq:Wellbore_PB}
    \end{eqnarray}
where $V(s)$ is the cross-sectional average fluid velocity at $s$, and $\alpha(s)$ the local inclination of the wellbore. Volume conservation under steady-state conditions implies $\pi R(s)^2V(s)=Q$ where $R(s)$ is the inner radius of the wellbore at $s$. In the previous expression $p_{\textrm{surf}}$ denotes the fluid pressure at the pumps on the surface, while $p_{in}$ is the fluid pressure at the inlet of the hydraulic fracture \textit{on the fracture side}, past the perforation and near-wellbore regions. The perforation friction loss coefficient can be estimated via engineering formula \citep{EcNo00,CrCo88}, while the near-wellbore friction index $\beta$ and corresponding coefficient is usually determined in-situ from step-down rate tests (see \cite{EcNo00,LeDe15c,LeDe15d}).
The viscous shear stress (friction) in the well is defined by:
\begin{equation}
    \tau_w =  f(R_e, k_r) \times \frac{1}{2}\rho^{F} V^2 
\end{equation}
where $f(R_e,k_r)$ is the steady-state Fanning friction factor which depends on the Reynolds number $R_e=\dfrac{\rho^F V (2R)}{\mu}$ and the "roughness" of the pipe $k_r=h_r/(2R)$ (with $h_r$ the roughness scale). It evolves as $f=16/R_e$ in laminar conditions ($R_e\le 2100$). In turbulent conditions ($R_e > 4000$), engineering formulas such as the ones proposed by \cite{Cole39,YaDo10} (or some approximations) reproduce satisfactorily experiments. 
    
In the wellbore power-balance \eqref{eq:Wellbore_PB}, we have assumed steady-state flow conditions and an incompressible fluid. Of course, at the start of the injection, prior to hydraulic fracture initiation, some energy is spent to pressurise the wellbore up to the fracture initiation pressure $\sigma_i >\sigma^o$. This energy term can be estimated as 
\begin{equation}
    E_{\textrm{pressurization}}=\frac{1}{2}(\sigma_i-\rho^F g Z_{in}) ^2 \times c_f \times \mathcal{V}_{wb} 
\end{equation}
where $c_f$ is the fluid compressibility, $\mathcal{V}_{wb}$ the wellbore volume and $Z_{in}$ the true vertical depth of the hydraulic fracture inlet.

\begin{table}[]
    \centering
    \begin{tabular}{c|c|c|c}
         $E^\prime$ & $K_{Ic}$ & $C_L$ & $\sigma^o$   \\
         35 GPa & 1.2 MPa.$\sqrt{\text{m}}$ & $10^{-6} \text{m}\sqrt{\text{s}}$   & 45 MPa  \\ 
         \hline & & & \\
        Wellbore radius & Wellbore length L & Injection depth $Z_{in}$ & $\mathcal{V}_{wb}$ \\
         8.2 cm   & 3 km & 2 km  & 84.4 $\text{m}^3$ \\
         \hline & & & \\
         $\rho^F$ & $\mu$ & $c_f$  & \\
         1000 & $5\,10^{-3}$ Pa.s & $4.54 \, 10^{-10}$ Pa$^{-1}$ & \\
         \hline & & & \\
         $Q$& $V_{in}$ &    Injection duration &  \\ 
         $0.053\,\text{m}^3/\text{s}$ &  2.5 m/s & 45 minutes & 
    \end{tabular}
    \caption{Parameters for a realistic single-entry hydraulic fracturing treatment: rock and in-situ properties, wellbore dimensions, fluid properties, injection rate (20 barrels per minutes), mean fluid velocity in the wellbore and injection duration.}
    \label{tab:parameters}
\end{table}

\subsection{Order of magnitude for an example - radial hydraulic fracture in a deviated well}
We can combine the wellbore power-balance \eqref{eq:Wellbore_PB} with the balance for the hydraulic fracture \eqref{eq:final_PB_2} to relate the actual surface energy input $\int_0^t p_{\textrm{surf}}(t') Q \text{d}t' $ with the different energy losses.  We do so here for the simple case of a single \textit{radial} hydraulic fracture in order to illustrate how to estimate the order of magnitude of the different terms. Of course, depending on the exact field conditions, the results will necessarily differ.
We set the rock and fluid properties as well as the in-situ conditions to "realistic" values (see Table \ref{tab:parameters}). For simplicity, we assume a wellbore of constant diameter (such that the kinetic losses are negligible) and focus on the case of so-called single-entry treatment (only one hydraulic fracture is being propagated). We neglect any near-wellbore tortuosity effects and estimate the perforation friction using the  \cite{CrCo88} formula for 6 perforations with a diameter of $2$ cm (and $C_p=0.5$). This results in a pressure drop of $1.57$MPa across the perforations. With the parameters chosen, the fluid flow is turbulent in the wellbore such that we estimate the wellbore friction loss there using the Colebrook formula from which we obtain $\tau_w \sim 2.76$Pa. It is worth noting that although the flow is turbulent in the wellbore, laminar conditions prevail in the hydraulic fracture after a very short transient after fracture initiation (see \cite{LeZi19} for a complete discussion). 

We estimate the order of magnitude of the different power terms for $45$ minutes of injection. For a radial geometry, with these parameters (Table \ref{tab:parameters}), the propagation remains in the viscosity-dominated regime: $t_{mk} \approx 350$ minutes, $t_{m\tilde{m}}\approx 1800$ minutes. 
We therefore use the viscosity-dominated scaling to estimate the different terms of the hydraulic fracture power-balance \eqref{eq:final_PB_2}. The buoyant term is negligible for this example and not reported. 

The order of magnitude of the energy spent for the total duration of the injection ultimately splits as follows.
\begin{itemize}
    \item Energy spent in overcoming the in-situ stress ($Q \sigma^o t$) 
    minus the energy gained by fluid gravity in the wellbore ($V_{in} \pi R^2 \rho^f g Z_{in} t$): 6438MJ - 2807 MJ = 3631MJ 
    \item Elastic stored energy in the medium: 101 MJ
    \item Energy dissipated in viscous flow inside the fracture: 202 MJ
    \item Energy dissipated in creating new fracture surfaces:  19 MJ
    \item Energy lost in fluid leak-off: 25.8 MJ
    \item Energy spent across the perforations: 225 MJ
    \item Energy spent in frictional loss along the wellbore: 38.5 MJ
    \item Energy spent pressurizing the wellbore up to fracture initiation (assuming $\sigma_i=1.2\sigma^o$): 22.18 MJ
\end{itemize}
which results in a total energy cost at the pump on the surface ($p_p Q t$) of 4.26 Giga Joules. This provides a characteristic surface pump pressure $p_p$ of the order of 30MPa. We provided here this simple example to highlight that most of the energy required for propagating a hydraulic fracture at depth is spent overcoming the minimum initial in-situ compressive stress $\sigma^o$ (85\% in this case). In practice, the knowledge of the fracturing depth is thus often sufficient to estimate the power required and therefore choose appropriate surface equipment. For the single entry treatment discussed, 1.6 Mega Watt is required on average, which can be easily provided by three 1100 horsepower fracturing pumps working at 50\% efficiency.

\section{Conclusions}

Recalling the different assumptions of the linear hydraulic fracture mechanics model, we have expressed the power balance of a propagating planar hydraulic fracture at depth. This energy approach allows to physically ground the different fracture propagation regimes originally delineated from scaling considerations of the strong form of the linear hydraulic fracture model. The ratio of the different power terms indeed controls how a hydraulic fracture propagates. We have illustrated the split of the different energies on several fracture propagation examples - some of them exhibiting a non-monotonic evolution of the different powers. Including energy losses in the fluid injection line, the order of magnitude of the power required in practice can be easily estimated.

The approach presented here can be extended to the case of multiple hydraulic fractures propagating simultaneously as well as to more complex fluid rheology. 
     

\appendix

\section{The energy release rate for hydraulically loaded cracks}\label{appendix:deriv_in_integral}

\begin{figure}[ht!]
\noindent \begin{centering}
\includegraphics[width=0.8\textwidth]{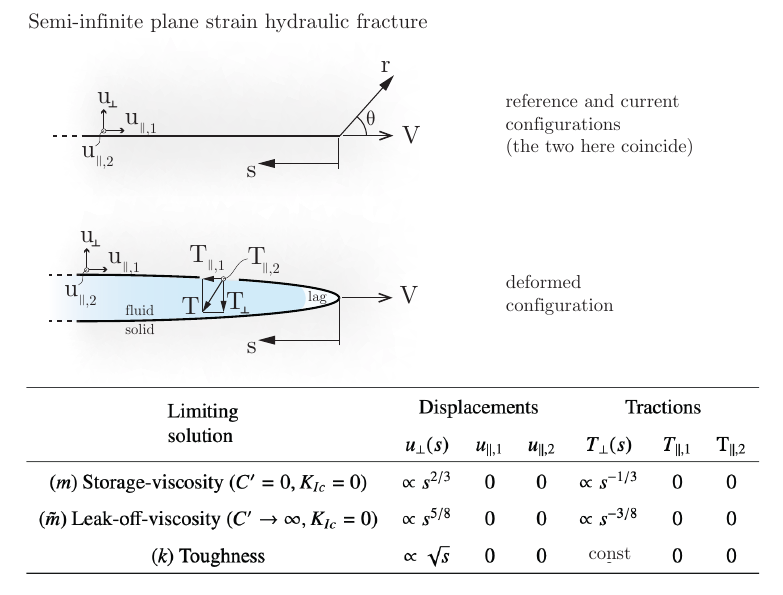}
\par\end{centering}
\caption{First-order asymptotic behaviours of the displacements $u_i$ and traction $T_i$ fields near the fracture tip of a semi-infinite, plane strain hydraulic fracture, as a function of the coordinate $s$ (equivalent to $\left(r,\theta=\pm \pi\right)$ ). This solution does not account for the hydraulically induced shear stress. $u_\perp$, and $u_{\parallel,1}$, $u_{\parallel,2}$ are respectively, the displacement perpendicular and parallel to the fracture plane. The tractions $T_i$ are expressed using the same convention.\label{fig:DT}}
\end{figure}

We demonstrate here the expression of the  energy release rate $\mathbbmsl{G}$~\eqref{eq:G_def}. 
We begin by using the principle of virtual works to express the elastic energy $\Psi^\textrm{S}$ as a surface integral over the entire boundary $\partial\Omega^\textrm{S}$. We assume zero initial stress for clarity - the principle of elastic superposition can always be used replacing $T_i$ with $T_i-T_i^o$. The total elastic potential energy $\phi=\Psi -\textrm{W}_{\textrm{ext}}$ \citep{Ri68}
\begin{equation}
\phi=\frac{1}{2}\int_{\partial\Omega^{\textrm{S}}}u_{i}T_{i}\mathrm{d}S-\int_{\partial\Omega^{\textrm{S}}_T}u_{i}T_{i}\mathrm{d}S,
\end{equation}
is further simplified by splitting the total boundary $\partial\Omega^\textrm{S}$ in the part where the external tractions are applied $\partial\Omega^\textrm{S}_T$ and the part where the displacements are imposed $\partial\Omega^\textrm{S}_u$ ($\textrm{d}\Omega^\textrm{S}=\textrm{d}\Omega^\textrm{S}_u \cup \textrm{d}\Omega^\textrm{S}_T$). The energy release rate $\mathbbmsl{G}$ then results from Eq.~\eqref{eq:G_2}
\begin{equation}
\mathbbmsl{G}=-\frac{1}{2}\frac{\mathrm{\mathrm{\partial}}}{\mathrm{\mathrm{\partial}}a}\left(\int_{\partial\Omega^{\textrm{S}}_u}u_{i}T_{i}\mathrm{d}S\right)+\frac{1}{2}\frac{\mathrm{\mathrm{\partial}}}{\mathrm{\mathrm{\partial}}a}\left(\int_{\partial\Omega^{\textrm{S}}_T}u_{i}T_{i}\mathrm{d}S\right).\label{eq:general_G}
\end{equation}
In case the crack faces belong to  $\partial\Omega^{\textrm{S}}_{T}$ and/or $\partial\Omega^{\textrm{S}}_{u}$, the derivative $\partial/\partial a$ of the corresponding integral in Eq.~\eqref{eq:general_G} can not be immediately moved inside of the integral sign because both the integrand and the limits of integration depends on the fracture area $a$. The expression of the derivative of an integral over a time-dependent region (see e.g. \cite{Rudn14}) proves useful in these situations. We apply it to the second integral on the right-hand side because we assume that the fracture surface belongs to $\partial\Omega^\textrm{S}_{T}$, and  obtain
\begin{equation}\label{appendix:Reynolds_theorem}
\frac{1}{2}\frac{\mathrm{\mathrm{\partial}}}{\mathrm{\mathrm{\partial}}a}\left(\int_{\partial\Omega^{\textrm{S}}_{T}}u_{i}T_{i}\mathrm{d}S\right)=\frac{1}{2}\int_{\partial\Omega^{\textrm{S}}_{T}}\frac{\mathrm{\mathrm{\partial}}}{\mathrm{\mathrm{\partial}}a}\left(u_{i}T_{i}\right)\mathrm{d}S+\frac{1}{2}\int_{\Gamma}\,u_{i}\,T_{i}\,\Tilde{v}_i\,n_i \, \mathrm{d}\ell,
\end{equation}
where both $n_i$, and $\Tilde{v}_i$ are quantities defined at each location $x_i$ of the fracture front $\Gamma$. $n_i$ is the $i-$component of the normal to the fracture front, and $\Tilde{v}_i$ is the $i-$component of the $\textit{pseudo-velocity}$ that is defined as
\[
\Tilde{v}_i=\frac{\partial x_i}{\partial a}, 
\]
and that is related to the actual fracture front velocity $v_i$, which is defined as
\[
v_i=\frac{\partial x_i}{\partial t}=\left.\frac{\partial x_i}{\partial a}\right|_t\frac{\partial a}{\partial t}. 
\]
We now discuss the last integral on the right-hand side of Eq.~\eqref{appendix:Reynolds_theorem}
\begin{equation}\label{appendix:front_int}
\int_{\Gamma}u_{i}\,T_{i}\,\Tilde{v}_i\,n_i\, \mathrm{d}\ell,
\end{equation}
where the work done to the system by the external traction at the fracture front $\left(u_iT_i\right)$ is the real discriminant of the value assumed by the integral~\eqref{appendix:front_int} because the $\textit{pseudo-velocity}$ $\Tilde{v}_i$ is in general non zero along the fracture front. The term $\left(u_iT_i\right)$ is evaluated by considering a region adjacent to the fracture front where the solution of the 3D hydraulic fracture problem reduces to the one of a plane strain semi-infinite hydraulic fracture moving at the local front speed $v_i$. Therefore, $\left(u_i T_i\right)$ in the integral \eqref{appendix:front_int} is expressed as 
\begin{equation}\label{appendix:lim}
u_iT_i=
\lim_{s\to0}{u_i\left(s\right)T_i\left(s\right)},
\end{equation}
where $s$ is a coordinate with its origin at the fracture front and increasing inside the fracture as represented in Fig.~\ref{fig:DT}. The value of the integral~\eqref{appendix:front_int} is identically zero regardless of the fact that we neglect, or not, the tangential stress exerted by the fluid onto the fracture surfaces. 

\paragraph{Negligible hydraulically induced shear stress}
The displacement field in a region around the propagating front of a plane strain steadily-moving semi-infinite hydraulic fracture is always influenced by the local fracture energy dissipation $G_\textrm{c} V$ ($G_\textrm{c}$ is the fracture energy, $V$ the local front velocity) which takes place at the front. In this case, the displacement field behaves asymptotically as the classical LEFM eigensolution indicated in Fig.~\eqref{fig:DT} as "($k$) Toughness" \citep{Wi52,GaDe00}. Two different solutions arise where there is no energy dissipation at the fracture front ($G_\textrm{c}=K_{\textrm{Ic}}=0$). These are the "storage-viscosity ($m$)", and the "leak-off viscosity ($\tilde{m}$)" solutions, see \cite{GaDe00,Leno95,GaAd11}. They correspond respectively to the case of zero and infinite Carter's leak off $C_L$. In each limiting solution ($k$, $m$, and $\tilde{m}$), the displacement and the traction fields are such that the external work is zero $u_i T_i=0$ at the fracture front. This can readily be verified by substituting in Eq.~\eqref{appendix:lim} the corresponding $u_i$ and $T_i$ behaviours reported in Fig.~\ref{fig:DT}.
\paragraph{Accounting for the hydraulically induced shear stress}
As discussed in \cite{WrPi17}, the LEFM asymptotic solution holds even when the fracture energy $G_\textrm{c}$ is zero. The first order behaviour of the opening displacement $u_\perp$ and pressure $T_\perp$ do not change with respect to the "$(k)$ Toughness" solution shown in Fig.~\eqref{fig:DT}. Instead, the displacement parallel to the fracture surface $u_{\parallel,1}$ is $\propto 0\cdot \sqrt{s}+\mathcal{O}\left( s \log (s)\right)$ and the shear tractions $T_{\parallel,1}$ are singular as $1/\sqrt{s}$ (see Table~\eqref{tab:app1}). As a consequence, the external work at the fracture front is also zero $u_i(s=0)T_i(s=0) = 0$.
\bgroup 
\def\arraystretch{1.5}
\begin{table}[h!]
\centering
\begin{tabular}{ccccccc} 
\hline
\multirow{2}{4em}{Limiting solution} & \multicolumn{3}{c}{Displacements} & \multicolumn{3}{c}{Tractions}\\
& $u_\perp(s)$ & $u_{\parallel,1}$ & $u_{\parallel,2}$ & $T_\perp(s)$ & $T_{\parallel,1}$ & $T_{\parallel,2}$ \\
\hline 
Toughness & $\propto \sqrt{s}$ & $\propto \left(0\cdot\sqrt{s}\right)+\mathcal{O}\left( s \log (s)\right)$ & $0$ & $\mathrm{const}$ & $\propto 1/\sqrt{s}$ & $0$ \\ 
\hline
\end{tabular}
\caption{First-order asymptotic behaviours of the displacements $u_i$ and traction $T_i$ fields near the fracture tip of a semi-infinite, plane strain hydraulic fracture, as a function of the coordinate $s$ (equivalent to $\left(r,\theta=\pm \pi\right)$). This solution accounts for the hydraulically induced shear stress. $u_\perp$, and $u_{\parallel,1}$, $u_{\parallel,2}$ are respectively, the displacement perpendicular and parallel to the fracture plane. The tractions $T_i$ are expressed using the same convention.}\label{tab:app1}
\label{table:1}
\end{table}
\egroup

\section{Thin-film lubrication \label{app:Lubrication}}

By using the thin film lubrication scaling \citep{szeri10}, we have reduced the  power balance for a fluid flowing between two surfaces to Eq.~\eqref{eq:Fluid_P_lubrication}
\[
\int_{\partial\Omega^\textrm{F}_T}{T_i\,\dot{u}_i\,\textrm{d}S}+\int_{\Omega^\textrm{F}_T} {\rho_\textrm{F}\,g_\alpha\,\dot{u}_\alpha\,\textrm{d}V}=\underset{\dispdot{\textrm{D}}_{\textrm{dev}-0}^\textrm{F}}{\underbrace{\int_{\Omega^{\textrm{F}}}\mu\left[\left(\frac{\partial\dot{u}_2}{\partial x_3}\right)^{2}+\left(\frac{\partial\dot{u}_{1}}{\partial x_3}\right)^{2}\right]\textrm{d}V}},\quad \alpha=1,2\quad i=1,2,3,
\]
where the external power on the left-hand side of the equation is entirely dissipated because of the velocity gradient across the thin film of thickness (direction $x_3$). Performing the same dimensional analysis on the Navier-Stokes equations, neglecting inertia, the terms of order $>\mathcal{O}\left(\epsilon\right)$, and assuming a Newtonian fluid leads to:
\begin{equation}
\frac{\partial p}{\partial x_1}=\underbrace{\mu\frac{\partial^2\dot{u}_1}{\partial x_3^2}}_{\partial \tau_{13}/\partial{x_3}}+\rho_\textrm{F} g_1,\quad\frac{\partial p}{\partial x_2}=\underbrace{\mu\frac{\partial^2\dot{u}_2}{\partial x_3^2}}_{\partial \tau_{23}/\partial{x_3}}+\rho_\textrm{F} g_2,\quad\frac{\partial p}{\partial x_3}=0.\label{eq:NS_3}
\end{equation}

As a first step to express the power balance as a function of the average fluid velocity $v_\alpha$ \eqref{eq:Average_u}
\[
v_\alpha=\frac{1}{w}\int_{\,-w/2}^{\,w/2}\dot{u}_\alpha\,\textrm{d}x_3\quad \alpha=1,2,
\]
across the film thickness $w$, we integrate the first two reduced Navier-Stokes equations~\eqref{eq:NS_3} along the fluid thickness  (direction $x_3$) assuming a no-slip boundary conditions at the fluid-solid interface: 
\begin{equation}
\dot{u}_\alpha^{\textrm{F}}=\dot{u}_\alpha^{\textrm{S}}\quad \alpha=1,2,\label{eq:no_slip_bc}
\end{equation}
The local solid velocity parallel to the fracture surface $\dot{u}_\alpha^{\textrm{S}}$, $\alpha=1,2$ can be neglected, implying that the fluid velocity at the interface is zero,
\begin{equation}
\dot{u}_\alpha^{\textrm{F}}\left(x_3=\pm w/2\right)=0\quad \alpha=1,2.
\end{equation}
 
As a result, we obtain the classical parabolic fluid velocity distribution across the fracture width  
\begin{equation}
\dot{u}_\alpha\left(x_1,x_2,x_3\right)=\frac{1}{2\mu}\left(\frac{\partial p}{\partial x_\alpha}-\rho_F g_\alpha\right)\left(x_3^2-\frac{w^2}{4}\right)\qquad \alpha=1,2,\label{eq:u_distr}
\end{equation}
and the expression of the average fluid velocity~\eqref{eq:Average_u} becomes 
\begin{equation}
v_\alpha\left(x_1,x_2\right)=-\frac{1}{12\mu}\left(\frac{\partial p_{net}}{\partial x_\alpha}+\frac{\partial \sigma_o}{\partial x_\alpha}-\rho_F g_\alpha\right)w^2\qquad \alpha=1,2,\label{eq:P_17-1}
\end{equation}
where we have expressed the fluid pressure $p$ as  $p=p_{net}+\sigma_o(x_1,x_2)$, with $\sigma_o(x_1,x_2)$ being the in-situ stress orthogonal to the fracture plane.

We then integrate by parts the right-hand side of the fluid power balance Eq.~\eqref{eq:Fluid_P_lubrication} along the $x_3$ direction obtaining
\begin{equation}
\displaystyle
\int_{\partial\Omega^\textrm{F}_T}{T_i\,\dot{u}_i\,\textrm{d}S}+\int_{\Omega^\textrm{F}_T} {\rho_\textrm{F}\,g_\alpha\,\dot{u}_\alpha\,\textrm{d}V}
=\int_\Sigma{\biggr{[}\dot{u}_\alpha\,\underset{\sigma_{3\alpha}}{\underbrace{\left(\mu\frac{\partial \dot{u}_\alpha}{\partial x_3}\right)}}\,\biggr{]}_{x_3=-w/2}^{x_3=w/2}\textrm{d}S}-\mu\int_{\Omega^\textrm{F}}\dot{u}_\alpha\frac{\partial^2\dot{u}_\alpha}{\partial x_3^2}\,\mathrm{d}V\quad \alpha=1,2\quad i=1,2,3,\label{eq:P_13}
\end{equation}
where $\Sigma$ is the middle surface of the fluid domain $\Omega^\textrm{F}$. 
The first integral on the right-hand side, in which the term in brackets is the shear stress $\sigma_{3\alpha}$, represents the power exchanged at the interface between the fluid and the solid and can be re-written as 
\[
\int_{\partial\Omega^\textrm{F}_T} \dot{u}_\alpha T_\alpha\textrm{d}S ,\quad \alpha=1,2
\]

The dissipated power in the fluid (last term on the right-hand side in Eq.~\eqref{eq:P_13}) by making use of the reduced Navier-Stokes equations  \eqref{eq:NS_3} can expressed as
\begin{equation}
-\mu\int_{\Omega^\textrm{F}}\dot{u}_\alpha\frac{\partial^2\dot{u}_\alpha}{\partial x_3^2}\,\mathrm{d}V=-\int_{\Omega^\textrm{F}}\dot{u}_\alpha\,\left(\frac{\partial p}{\partial x_\alpha}-\rho_\textrm{F}\,g_\alpha\right)\textrm{d}V\qquad \alpha=1,2\quad i=1,2,3,
\label{eq:P_15}
\end{equation}
The 3rd of the reduced Navier-Stokes equations \eqref{eq:NS_3} implies that the fluid pressure is constant across the fluid thickness, in other words $p=p(x_1,x_2)$. This allows to integrate the dissipation term (right-hand side of Eq.~\eqref{eq:P_15}) across the thickness (direction $x_3$), to obtain:
\begin{equation}
-\mu\int_{\Omega^\textrm{F}}\dot{u}_\alpha\frac{\partial^2\dot{u}_\alpha}{\partial x_3^2}\,\mathrm{d}V= 12\mu\int_\Sigma\frac{v_\alpha^2}{w}\,\textrm{d}S\qquad \alpha=1,2\quad i=1,2,3.
\label{eq:P_18}
\end{equation}

\newpage

\bibliographystyle{agsm}
\bibliography{EB_Bibliography,BibDB-Complete}

\end{document}